\documentstyle[11pt]{article} \def\be{\begin{equation}}
\def\ee{\end{equation}} \def\ba{\begin{eqnarray}}
\def\ea{\end{eqnarray}} \textwidth16.0cm \textheight22.0cm
 \date{}
 
\begin{document}
\begin{titlepage} \begin{flushright}  HD--THEP--94--3
\end{flushright} \quad\\ \vspace{2cm} \begin{center} {\bf\LARGE Two
loop results from one loop computations}\\ \bigskip {\bf\LARGE and
non perturbative solutions of exact evolution equations}\\
\vspace{1cm} T. Papenbrock and C. Wetterich\\ \bigskip {\em
Institut f\"ur Theoretische Physik, Universit\"at Heidelberg,}\\
{\em Philosophenweg 16, 69120 Heidelberg}\\ \vspace{3cm} {\bf
Abstract:} \\ \parbox[t]{\textwidth}{ A nonperturbative method is
proposed for the approximative solution of the exact evolution
equation which describes the scale dependence of the effective
average action. It consists of a combination of exact evolution
equations for independent couplings with renormalization group
improved one loop expressions of secondary couplings. Our  method
is illustrated by an example: We compute the \(\beta\)-function of
the quartic coupling \(\lambda\) of an \(O(N)\) symmetric scalar
field theory to order \(\lambda^3\) as well as the anomalous
dimension to order \(\lambda^2\) using only one loop expressions
and find agreement with the two loop perturbation theory. We also
treat the case of very strong coupling and confirm the existence of
a "triviality bound".} \end{center} \end{titlepage} \newpage
\section{Introduction} Nonperturbative field theoretical methods
formulated in continuous space and allowing practical calculations
are relatively rare. Among the best known figure the
Schwinger-Dyson equations \cite{1} wich constitute a system of
exact equations for the 1PI vertices. The momentum integrals
appearing in these equations contain, however, not only 1PI
vertices but also "bare" couplings. Their solution requires an
understanding of physics at very different length scales, englobing
the "short-distance physics" and the "long-distance physics" whose
scale is characterized by the external momenta of the 1PI green
functions. The necessity to integrate over a large momentum range
(within which the relevant physics may vary) makes it often hard to
find approximate methods which go beyond perturbation theory in
some small dimensionless coupling.\\ The block spin approach
\cite{2} in lattice theories seems to be an ideal solution for this
type of problem: Only the understanding of the physics within a
small momentum range is needed in order to make the transition from
one length scale to a larger one. The effective average action
\(\Gamma_k\) \cite{3}-\cite{5} formulates the block spin concept in
continuous space. The average action is expressed as a functional
integral with a constraint which ensures that only quantum
fluctuations with momenta \(q^2>k^2\) are effectively included.
Various field theoretical methods can be used for an approximate
solution of this functional integral, as in particular, a
renormalization group improved saddle point approximation
\cite{3}-\cite{5}. This method is nonperturbative in the sense that
no small dimensionless coupling is required. Nonperturbative
phenomena in two and three dimensionsal scalar theories \cite{4},
as well as in four dimensional scalar theories at nonvanishing
temperature \cite{6} have been described successfully by this
method.\\ An infinitesimal analogue of the Schwinger-Dyson
equations for the effective average action describes the change of
\(\Gamma_k\) with varying scale \(k\). This ensures that only a
small momentum range \(q^2\approx k^2\) must be contolled. The
change of the \(k\)-dependent 1PI vertices characterizing
\(\Gamma_k\) can now be expressed in terms of such vertices only.
Indeed, an exact evolution equation for the scale dependence of the
effective average action has been proposed recently \cite{7}. It
reads for scalar fields \(\varphi\) \begin{equation} \label{master}
\frac{\partial}{\partial{t}}\Gamma_{k}[\varphi]=
\frac{1}{2}{T}{r}\left\{(\Gamma_{k}^{(2)}[\varphi]+{R}_{k})^{-1}
\frac{\partial}{\partial{t}}
{R}_{k}\right\}. \end{equation} Here \({t} =
\ln\frac{{k}}{\Lambda}\) where \(\Lambda^{-1}\) is a suitable short
distance scale and \(\Gamma_{k}^{(2)}\) is the exact \(k\)
dependent inverse propagator as given by the second functional
derivative of \(\Gamma_k\) with respect to the fields \(\varphi\).
The trace indicates a summation over internal indices and a
momentum integration \footnote{We use \(Tr = \sum_q\sum_a =
(2\pi)^{-d}\Omega\int{d}^{d}q\sum_a\) where \(\Omega\) is the
volume of space-time. For convenience we often use
\(\Gamma_k^{(2)}\) for the second functional derivative devided by
\(\Omega\).}. The function \({R}_{k}\) contains details of the
averaging procedure and causes the integral to be infrared
convergent. Ultraviolet finiteness is guaranted by an appropriate
behaviour of \(\frac{\partial{R_k}}{\partial{t}}\). We choose
\begin{equation} \label{two} R_k =
\frac{Z_{k}{q^2}f_k^2(q)}{1-f_k^2(q)} \end{equation} with
\begin{equation} \label{ffunc} f_k^2(q) = \exp{-\frac{q^2}{k^2}}.
\end{equation} We note that for \(q^2 \ll k^2\) the infrared cutoff
\(R_k\approx Z_{k}k^2\) acts like an additional mass term whereas
its influence on modes with \(q^2 \gg k^2\) is suppressed
exponentially. Although eq. (\ref{master}) has the form of a
renormalization group improved one loop equation (and has actually
first been proposed in this context \cite{4}) it involves no
approximation. It has a simple graphical representation (fig.1) as
a loop expression in terms of 1PI-vertices only. Equation
(\ref{master}) can be related  to earlier versions of "exact
renormalization group equations" \cite{8} by an appropriate
Legendre transform \cite{9}. For \(k\to 0\) all quantum
fluctuations are "integrated out" and the effective average action
becomes the generating functional for the 1PI green functions (the
usual "effective action") in this limit. This distinguishes
\(\Gamma_k\) from the "effective action with variable ultraviolet
cutoff" whose dependence on the cutoff is described by the "exact
renormalization group equation" \cite{8} and which becomes a
relatively complicated object in the limit \(k\to 0\) \cite{10}.
The simple form of \(\Gamma_k\), which can be described, for
example, in terms of a scalar potential, kinetic term and higher
derivative terms, makes it a reasonable starting point for the
development of a nonperturbative method.\\ A nonperturbative exact
evolution equation (\ref{master}) is not yet a nonperturbative
method. In order to extract physics one needs in addition a
systematic way of solving this equation for \(k\to 0\), with
initial conditions specified by the "bare action"
\(\Gamma_\Lambda\) at some short distance scale \(\Lambda^{-1}\).
As it stands, eq. (\ref{master}) is a partial differential equation
for a function \(\Gamma_k\) depending on infinitely many variables
- for example the Fourier modes \(\varphi(q)\) of the scalar field
and \(k\). This is, in general, impossible to solve. Equivalently,
we may expand \(\Gamma_k\) in terms of invariants with respect to
the symmetries of the theory. Eq. (\ref{master}) is then
transformed into an infinite system of coupled nonlinear
differential equations for infintely many couplings (for example
1PI Green functions). Without a systematic approximation method
which reduces this infinite system to a finite system (which can
then be solved by numerical or analytical methods) not much
practical progress is made compared to the simple renormalization
group improvement discussed in ref. \cite{4}. From a more
systematic point of view a nonperturbative method should be able to
compute physical quantities and to give an estimate of the error
even for situations where perturbation theory fails. Our paper
contains a proposal for such a systematic nonperturbative method.\\
The basic problem we are confronted with can be seen by taking
functional derivatives of eq. (\ref{master}) with respect to
\(\varphi\). We need \(\Gamma^{(2)}_k\) on the r.h.s. and may
obtain an evolution equation for this quantity by taking the second
functional derivative of eq. (\ref{master}). The corresponding flow
equation expresses \(\frac{\partial{\Gamma^{(2)}_k}}{\partial{t}}\)
in terms of \(\Gamma^{(2)}_k\), \(\Gamma^{(3)}_k\)\ and
\(\Gamma^{(4)}_k\). This feature proliferates to higher Green
functions: The evolution equation for the n-point function
\(\Gamma^{(n)}_k\) involves \(\Gamma^{(n+1)}_k\) and
\(\Gamma^{(n+2)}_k\) and the system never closes. For \(n\)
sufficiently large the couplings \(\Gamma^{(n)}_k\) are "irrelevant
couplings". They are usually small in a perturbative context since
they are proportional to appropriate powers of a small coupling.
For nonperturbative problems no such argument is available and we
have to find a way to estimate \(\Gamma^{(n+1)}_k\) and
\(\Gamma^{(n+2)}_k\) in order to solve the evolution equation for
\(\Gamma^{(n)}_k\). The main point for a proposal for a systematic
nonperturbative method for scalar field theories consists in the
calculation of \(\Gamma^{(n)}_k\) for \(n\) sufficiently large by a
renormalization group improved one loop approximation. In contrast
to the exact expression for
\(\frac{\partial{\Gamma^{(n)}_k}}{\partial{t}}\) which can be
derived from (\ref{master}) the one loop expression for
\(\Gamma^{(n)}_k\) will be approximative. It only involves 1PI
vertices \(\Gamma^{(m)}_k\) with \(m\leq n\). For high enough \(n\)
the momentum integral in the loop is both ultraviolet and infrared
finite and dominated by a small momentum range \(q^2\approx k^2\)
\footnote{For smaller n this holds only for
\(\frac{\partial{\Gamma^{(n)}_k}}{\partial{t}}\), not for the one
loop expression of \(\Gamma^{(n)}_k\) itself.}. We therefore expect
this approximation to be quite accurate if the involved vertices
$\Gamma_k^{(m)}$ do not depend to strongly on $k$. Details of our
method and a way of estimating errors will be indicated in the next
section.\\ We also want to test our proposal for a new
nonperturbative method for some problem where exact analytical
results are known. For this reason we compute in the present paper
the \(\beta\)-function for the quartic coupling \(\lambda\) of a
\(O(N)\)-symmetric scalar theory in four dimensions.  In section 3
we derive in a short way the general equations. In sections 4 to 6
we calculate the  beta function of the quartic coupling step by
step increasing the number of included couplings. We keep all terms
in order \(\lambda^3\) and find agreement with the standard two
loop result. We emphasize, however, that we never perform a two
loop calculation but only solve the exact evolution equation
(\ref{master}). We also do not employ here an iterative solution
which would be equivalent to a loop expansion: All momentum
integrals appearing in our calculation have the form of one loop
integrals. In our approach the two loop result obtaines from an
improved one loop calculation! Although this result is remarkable
by itself it should be seen here as an illustration and check of
our nonperturbative method. The real power of this method will
appear once applied to truly nonperturbative problems. A precision
calculation of critical exponents in the three dimensional theory
is in progress and results will be reported elsewhere. As discussed
in more detail in the next section the proposed method has its
limitations for a very fast running of the couplings. We will deal
explicitly with such a situation in sect. 7, where we discuss the
evolution equations for very large values of the quartic scalar
coupling. The results of sect. 7 confirm the existence of a
triviality bound in the context of continuum field theory.
\setcounter{equation}{0} \section{A nonperturbative method for
scalar field theories} The evolution equation (\ref{master}) has a
close resemblence with a one loop expression: \begin{eqnarray}
\label{master2} \frac{\partial}{\partial{t}}\Gamma_{k} &=&
\frac{1}{2}{T}{r}\left\{\left(\Gamma_{k}^{(2)}+{R}_{k}\right)^{-1}
\frac{\partial}{\partial{t}}{R}_{k}\right\}\nonumber\\
&=&\frac{1}{2}{T}{r}\left\{\frac{\partial}{\partial{t}}\ln
\left(\Gamma_{k}^{(2)}+{R}_{k}\right)\right\}
-
\frac{1}{2}{T}{r}\left\{\left(\Gamma_{k}^{(2)}+{R}_{k}\right)^{-1}
\frac{\partial}{\partial{t}}{\Gamma}_{k}^{(2)}\right\}.
\end{eqnarray} Let us define the "renormalization group improved
one loop contribution" \(\Gamma_k^{(L)}\) by \begin{eqnarray}
\label{olcon} \Gamma_k^{(L)} &=&
\frac{1}{2}Tr\ln\frac{\Gamma_k^{(2)}+R_k}{\Gamma_\Lambda^{(2)}+
R_\Lambda}\nonumber\\
&=&
\frac{1}{2}\ln{Det}\left(\Gamma_k^{(2)}+R_k\right)-\frac{1}{2}
\ln{Det}\left(\Gamma_\Lambda^{(2)}+R_\Lambda\right)
\end{eqnarray} and identify \(\Gamma_\Lambda\) with the "classical
action" \(S\). Then we can write \begin{equation} \Gamma_k = S +
\Gamma_k^{(L)} + \Delta\Gamma_k \end{equation} where the
"correction term" \(\Delta\Gamma_k\) vanishes in the limit where
the last term in eq. (\ref{master2}) proportional to
\(\frac{\partial}{\partial{t}}\Gamma_k^{(2)}\) can be neglected. We
emphasize that \(\Gamma_k^{(L)}\) is already an ultraviolet
regulated one loop expression and vanishes for \(k\to\Lambda\).
This new form of an ultraviolet regulator simply subtracts the
contribution from quantum fluctuations with momenta \(q^2 >
{\Lambda}^2\) in a way very similar to the infrared regularization
discussed above. It can be easily generalized beyond the scalar
theory discussed in the present paper, for example to chiral
fermions \cite{11} or gauge theories \cite{12}, and may therefore
be useful for standard one loop calculations in a wider context
than the present paper. The main advantage of this regulator is the
conservation of all symmetries which are conserved by the quadratic
form \begin{equation} \Delta{S}_k =
\frac{1}{2}\Omega\sum_q\varphi^{+}(q){R}_{k}(q)\varphi(q)
\end{equation} which provides the infrared cutoff in a functional
integral representation of \(\Gamma_k\) \cite{7}. \\ The
disadvantage of the immediate use of \(\Gamma_k^{(L)}\) is linked
to the fact that for most theories \(\Gamma_k^{(L)}\) depends
strongly on \(\Lambda\) for \(\frac{\Lambda}{k}\to\infty\). This
means that the "ultraviolet properties" encoded in
\(\Gamma_\Lambda^{(2)}\) have a strong influence on
\(\Gamma_k^{(L)}\), in contradiction to our strategy to describe
physics at the scale \(k\) only by properties of \(\Gamma_k\). For
this reason the renormalization group improved one loop
approximation was used in earlier work \cite{4} only for
\(\frac{\partial}{\partial{t}}\Gamma_k\) rather than for
\(\Gamma_k\), since \(\frac{\partial}{\partial{t}}\Gamma_k\)
becomes independent of \(\Lambda\) for \(\Lambda\to\infty\). The
situation improves greatly, however, if instead of \(\Gamma_k\) we
consider derivatives of \(\Gamma_k\) with respect to the fields,
i.e. appropriate 1PI Green functions with \(n\) external fields.
For \(n\) sufficiently large the corresponding one loop expression
\(\Gamma_k^{(n)(L)}\) becomes independent of \(\Lambda\) in the
limit \(\frac{\Lambda}{k}\to\infty\) \footnote{Strictly speaking
this holds sometimes only for an approximation of
\(\Gamma_k^{(2)}\) as discussed below.} and the momentum integral
(\ref{olcon}) is completely dominated by a narrow range
\(q^2\approx k^2\). For \(k\) small compared to \(\Lambda\) we can
then take the limit \(\Lambda\to\infty\) in order to obtain an
expression involving only \(\Gamma_k\). From the evolution equation
\begin{eqnarray} \label{noname}
\frac{\partial}{\partial{t}}\Gamma_k^{(n)} &=& \Omega^{-1}
\frac{\partial}{\partial{t}}\frac{\delta^n\Gamma_k}{\delta\varphi_1
\cdots\delta\varphi_n}
\nonumber\\
&=&\frac{1}{2}{T}{r}\left\{\frac{\partial}{\partial{t}}{R}_{k}
\Omega^{-1}\frac{\delta^n}{\delta\varphi_1\cdots\delta\varphi_n}
\left(\Gamma_{k}^{(2)}+{R}_{k}\right)^{-1}\right\}\nonumber\\
&=&\frac{\partial}{\partial{t}}\frac{1}{2}{T}{r}\left\{\Omega^{-1}
\frac{\delta^n}{\delta\varphi_1\cdots\delta\varphi_n}\ln
\left(\Gamma_{k}^{(2)}+{R}_{k}\right)\right\}\nonumber\\
&&-\frac{1}{2}{T}{r}\Omega^{-1}\frac{\delta^n}{\delta\varphi_1
\cdots\delta\varphi_n}\left\{\left(\Gamma_{k}^{(2)}+
{R}_{k}\right)^{-1}
\frac{\partial}{\partial{t}}{\Gamma}_{k}^{(2)}\right\}
\end{eqnarray}  we obtain \begin{eqnarray} \label{split}
\Gamma_k^{(n)} &=& \Gamma_k^{(n)(L)} +
\Delta\Gamma_k^{(n)}\nonumber\\
\Gamma_k^{(n)(L)}&=&\frac{1}{2}{T}{r}\left\{\Omega^{-1}
\frac{\delta^n}{\delta\varphi_1\cdots\delta\varphi_n}\ln
\left(\Gamma_{k}^{(2)}+{R}_{k}\right)\right\}.
\end{eqnarray} We observe that for \(n\) sufficiently high
\(\Gamma_\Lambda^{(n)}\) corresponds to an irrelevant operator such
that \(\Gamma_\Lambda^{(n)}\) can be neglected compared to
\(\Gamma_k^{(n)}\) for \(\frac{\Lambda}{k}\to\infty\). The
ultraviolet finitness (for \(\Lambda\to\infty\)) of the momentum
integrals in eq. (\ref{split}) serves as a direct test for wich
\(\Gamma_k^{(n)}\) this procedure of taking \(\Lambda\to\infty\) is
self-consistent.\\ Our strategy for a precision estimate of
\(\Gamma_k\) is composed of two parts. First we compute a set of
exact evolution equations for a set of 1PI Green functions
\(\Gamma_k^{\{A\}}\) by taking the appropriate functional
derivatives of eq. (\ref{master}). This set denoted by \(\{A\}\)
must contain at least all relevant or marginal couplings of the
theory but can, in general, be larger. We will call these couplings
the "independent" couplings since they are all treated on equal
footing for a solution of the evolution equations. The r.h.s. of
these evolution equations will depend not only on
\(\Gamma_k^{\{A\}}\) but also on some additional irrelevant
couplings \(\Gamma_k^{\{B\}}\). As a second step the "secondary"
couplings in the set \(\{B\}\) are computed by the renormalization
group improved one loop approximation (\ref{split}). Here it is
important that \(\Gamma_k^{(2)}\) on the r.h.s. of eq.
(\ref{split}) is computed for a truncation of \(\Gamma_k\)
corresponding to a subset of the 1PI vertices contained in
\(\{A\}\). This subset always contains the relevant and marginal
couplings. More generally, possible choices of subsets of \(\{A\}\)
are determined by the requirement of ultraviolet finiteness of eq.
(\ref{split}). This approximation of \(\Gamma_k^{\{B\}}\) is then
inserted in the evolution equation for \(\Gamma_k^{\{A\}}\). The
system is thereby closed and the evolution equation (\ref{master})
reduced to a finite set of coupled nonlinear differential equations
for the independent couplings \(\Gamma_k^{\{A\}}\) which can be
solved by numerical or analytical methods.\\ Furthermore, we may
consider \(\Gamma_k^{\{B\}(L)}\) as the zeroth order of an
iterative procedure for the determination of the secondary
couplings $\Gamma_k^{\{B\}}$. In fact, we may compute
\begin{equation} \label{error}
\frac{\partial}{\partial{t}}\Delta\Gamma_k^{(n)} =
-\frac{1}{2}{T}{r}\Omega^{-1}\frac{\delta^n}{\delta\varphi_1\cdots
\delta\varphi_n}\left\{\left(\Gamma_{k}^{(2)}+{R}_{k}\right)^{-1}
\frac{\partial}{\partial{t}}{\Gamma}_{k}^{(2)}\right\}
\end{equation} approximately by inserting on the r.h.s. the
couplings \(\Gamma_k^{\{A\}}\) and \(\Gamma_k^{\{B\}(L)}\).
Comparing this expression with
\(\frac{\partial}{\partial{t}}\Gamma_k^{(n)(L)}\) computed from eq.
(\ref{split}) gives an estimate of the error for the
\(\beta\)-function for couplings \(\Gamma_k^{\{B\}}\). Such an
inherent estimate of the error is particularly important for
nonpertubative applications of the exact evolution equations where
it is not easy to check the accuracy of the results otherwise. On
the other hand the improved \(\beta\)-function
\(\frac{\partial}{\partial{t}}\Gamma_k^{(n)(L)}+
\frac{\partial}{\partial{t}}\Delta\Gamma_k\)
can be used for an even more precise estimate of \(\Gamma_k\) by
solving the evolution equations for the combinet set
\(\Gamma_k^{\{A\}}\), \(\Gamma_k^{\{B\}}\). Further checks and
improvements can be obtained by systematically enlarging the set of
independent couplings \(\Gamma_k^{\{A\}}\) for which exact
evolution equations are used.\\ An obvious limitation of our method
arises for situations where the contribution from the term
$\frac{\partial}{\partial t}\Gamma_k^{(2)}$ in eq. (\ref{noname})
becomes comparable or even larger than the corresponding
contribution from $\frac{\partial}{\partial t}R_k$. The use of
renormalization group improved one loop results for the secondary
couplings gives in this case not the correct evolution equations
for these couplings. To get an idea when this happens in scalar
field theories we write
$\Gamma_k^{(2)}=Z_kq^2+c\lambda(k)\varphi^2+\ldots$. The first term
gives a contribution proportional to the anomalous dimension. The
second contributes substantially only for a fast running of
$\lambda(k)$. This typically happens for very strong $\lambda$. We
will discuss the $\beta$-function for $\lambda$ for the case of
strong quartic coupling in sect. 7 and sketch there alternative
estimates for the secondary couplings.\\ \setcounter{equation}{0}
\section{Exact evolution equation for the average potential} We
consider an \(O(N)\)-symmetric scalar field theory in arbitrary
dimension \(d\). We expand the effective average action in terms of
the average potential, kinetic term and higher derivatives.
\begin{eqnarray} \label{expan}
\Gamma_{k}[\varphi]=\int{d}^{d}{x}\bigg\{{U}_{k}(\rho)+\frac{1}{2}
\partial_\mu\varphi^{a}:{Z}_{k}(\rho,-\partial^2):
\partial^\mu\varphi_{a}
\nonumber\\
+\frac{1}{4}\partial^\mu\rho:{Y}_{k}(\rho,-\partial^2):
\partial_\mu\rho
+\ldots\bigg\} \end{eqnarray} where \(\rho =
\frac{1}{2}\varphi^a\varphi_a\) and normal ordering indicates that
the derivative only acts to the right. Eq. (\ref{expan}) is the
most general effective action which is $O(N)$-symmetric and
contains no more than two derivatives. The exact evolution equation
for the effective average potential can be obtained from
(\ref{master}) by expanding around a constant background field
\footnote{The neglected higher derivatives do not contribute to
\(\frac{\partial{U_k}}{\partial{t}}\).}. One finds \begin{equation}
\label{expot}
\frac{\partial}{\partial{t}}{U}_{k}(\rho)=\frac{1}{2}\int
\frac{{d}^{d}{q}}{(2\pi)^{d}}\frac{\partial}{\partial{t}}
{R}_{k}({q})\left(\frac{{N}-1}{{M}_0}+\frac{1}{{M}_1}\right)
\end{equation}  with \begin{eqnarray}
{M}_0(\rho,{q}^2)&=&{Z}_{k}(\rho,{q}^2){q}^2+{R}_{k}({q})+
{U}'_{k}(\rho)
\nonumber\\
{M}_1(\rho,{q}^2)&=&\tilde{Z}_{k}(\rho,{q}^2){q}^2+{R}_{k}({q})+
{U}'_{k}(\rho)+2\rho{U}''_{k}(\rho)
\end{eqnarray} and \begin{equation}
\tilde{Z}_k(\rho,{q}^2)={Z}_{k}(\rho,{q}^2)+\rho{Y}_{k}(\rho,{q}^2).
\end{equation} Equation (\ref{expot}) is a nonlinear partial
differential equation for a function of two variables \(t\) and
\(\rho\). One needs, however, the wave function renormalizations
\({Z}_{k}(\rho,{q}^2)\) and  \(\tilde{Z}_k(\rho,{q}^2)\) which must
be estimated, in turn, by a solution of evolution equations or by
alternative methods. We are mainly interested in the behaviour near
the phase transition, i.e. the massless theory or the theory with
scalar mass much smaller than the cutoff scale \(\Lambda\). Then
the evolution of \(U_k\) is mainly described \cite{3}, \cite{4} by
the regime with spontaneous symmetry breaking where the minimum of
the potential occurs for \(\rho_0 > 0\) with a \(k\) dependent
location of the minimum \(\rho_0(k)\) determined by \(U'_k(\rho_0)
= 0\). (Primes denote derivatives with respect to \(\rho\).) It is
convenient to expand the potential in powers of \(\rho - \rho_0\).
This transforms eq. (\ref{expot}) into an infinite system of
ordinary differential equations for the infinitely many couplings
\(U^{(n)}_k(\rho_0)\). In particular, the first and second
derivative of eq. (\ref{expot}) with respect to \(\rho\) yields
\begin{eqnarray} \label{three}
\frac{\partial}{\partial{t}}U'_k(\rho)&=&v_d(N-1)k^{d-2}
U''_k(\rho)L_{1,0}^d(w_1)
+ v_d(N-1)k^dZ'_k(\rho)L_{1,0}^{d+2}(w_1) \nonumber\\
&&+v_dk^{d-2}\left(3U''_k(\rho)+2U'''_k(\rho)\rho\right)
{L}_{0,1}^d(w_2)+
v_dk^d\tilde{Z}'_k(\rho){L}_{0,1}^{d+2}(w_2),  \end{eqnarray}
\begin{eqnarray} \label{four}
\frac{\partial}{\partial{t}}U''_k(\rho)&=&
-v_d(N-1)k^{d-4}\left(U''_k(\rho)\right)^2L_{2,0}^d(w_1)
-v_dk^{d-4}\left(3U''_k(\rho)+2U'''_k(\rho)\rho\right)^2
{L}_{0,2}^d(w_2)
\nonumber\\
&&-2v_d(N-1)k^{d-2}U''_k(\rho)Z'_k(\rho)L_{2,0}^{d+2}(w_1)
-2v_dk^{d-2}\left(3U''_k(\rho)+2U'''_k(\rho)\rho\right)
\tilde{Z}'_k(\rho){L}_{0,2}^{d+2}(w_2)\nonumber\\
&&-v_d(N-1)k^d\langle\left(Z'_k(\rho)\right)^2
\rangle{L}_{2,0}^{d+4}(w_1)
-v_dk^d\langle\left(\tilde{Z}'_k(\rho)\right)^2
\rangle{L}_{0,2}^{d+4}(w_2)\nonumber\\
 &&+v_{d}(N-1)k^{d-2}U'''_k(\rho)L_{1,0}^{d}(w_1)
+v_dk^{d-2}\left(5U'''_k(\rho)+2U^{(4)}_k(\rho)\rho\right)
{L}_{0,1}^d(w_2)
\nonumber\\ &&+v_d(N-1)k^dZ''_k(\rho)L_{1,0}^{d+2}(w_1) +
v_dk^d\tilde{Z}''_k(\rho){L}_{0,1}^{d+2}(w_2).   \end{eqnarray}
Here  \begin{equation} v_d^{-1} = 2^{d+1}\pi^{d\over{2}}\Gamma(d/2)
\end{equation} and we have used the variables \begin{eqnarray}
\label{double_u} w_1&=&U'_k(\rho), \nonumber\\
w_2&=&U'_k(\rho)+2U''_k(\rho)\rho.  \end{eqnarray} In terms of the
shorthands \begin{eqnarray} \label{exprops} P =
Z_k(\rho,x)x+R_k(x), \nonumber\\ \tilde{P} =
\tilde{Z}_k(\rho,x)x+R_k(x)  \end{eqnarray} the dimensionless
momentum integrals \(L_{n_1,n_2}^d\) are given by \begin{equation}
\label{five} L_{n_1,n_2}^d(w_1,w_2) =
k^{2(n_1+n_2)-d}\int_0^{\infty}dxx^{\frac{d}{2}-1}
\frac{\partial}{\partial{t}}\left\{(P+w_1)^{-n_1}
(\tilde{P}+w_2)^{-n_2}\right\}
\end{equation} where \(\partial/\partial{t}\) acts on the r.h.s.
only on \(R_k\) and \(x=q^2\). The functions \(Z'_k(\rho)\) and
\(\langle(Z'_k(\rho))^2\rangle\) correspond to appropriate moments
\begin{eqnarray} \label{moma} \left(Z'_k(\rho)\right)^d_{n_1,n_2}
&=&
\frac{\int\limits_0^\infty{dx}x^{\frac{d}{2}-1}Z'_k(\rho,x)
\frac{\partial}{\partial{t}}\left\{(P+w_1)^{-n_1}
(\tilde{P}+w_2)^{-n_2}\right\}}{\int\limits_0^\infty{dx}
x^{\frac{d}{2}-1}\frac{\partial}{\partial{t}}\left\{(P+w_1)^{-n_1}
(\tilde{P}+w_2)^{-n_2}\right\}},
\\ \label{momb} \langle(Z'_k(\rho))^2\rangle^d_{n_1,n_2} &=&
\frac{\int\limits_0^\infty{dx}x^{\frac{d}{2}-1}
\left(Z'_k(\rho,x)\right)^2\frac{\partial}{\partial{t}}
\left\{(P+w_1)^{-n_1}(\tilde{P}+w_2)^{-n_2}\right\}}{\int
\limits_0^\infty{dx}x^{\frac{d}{2}-1}\frac{\partial}{\partial{t}}
\left\{(P+w_1)^{-n_1}(\tilde{P}+w_2)^{-n_2}\right\}}
 \end{eqnarray} and similar for \(Z''_k(\rho)\),
\(\tilde{Z}_k(\rho)\) etc. We have omitted for simplicity of
notation the labels \(n_1\),\(n_2\) and \(d\) as well as the
dependence of these moments on \(w_1\) and \(w_2\). Those can be
infered easily from accompanying functions
\(L_{n_1,n_2}^d(w_1,w_2)\).\\ The minimum \(\rho_0(k)\) of the
potential is defined by the condition \begin{equation} \label{six}
U'_k(\rho_0) = 0. \end{equation} We are interested in the running
of the minimum \(\rho_0\). For notational simplicity we denote
\(U^{(n)}_k(\rho_0)\), \(Z^{(n)}_k(\rho_0)\),
\(\tilde{Z}^{(n)}_k(\rho_0)\) by \(U^{(n)}\), \(Z^{(n)}\),
\(\tilde{Z}^{(n)}\) respectively and understood that \(Z'^2\) means
the moment (\ref{momb}) evaluated at \(\rho=\rho_0\). Taking the
total derivative of (\ref{six}) we get with (\ref{three})
\begin{eqnarray} \label{seven} \bar{\delta} \equiv
\frac{d}{dt}\rho_0 &=&
-\frac{\partial{U'_k(\rho_0)}}{\partial{t}}\bar{\lambda}^{-1}
\nonumber\\
&=& -v_d(N-1)k^{d-2}L_{1,0}^{d}(0) -
v_d(N-1)k^{d}L_{1,0}^{d+2}(0){Z'}/{\bar{\lambda}} \nonumber\\
&&-v_{d}k^{d-2}\left(3+2U'''\rho_0/{\bar{\lambda}}\right)
{L}_{0,1}^{d}
(2\bar{\lambda}\rho_0)
-v_{d}k^{d}{L}_{0,1}^{d+2}(2\bar{\lambda}\rho_0)
\tilde{Z}'/{\bar{\lambda}}.
\end{eqnarray} We  also define the quartic coupling \(\bar{\lambda}
\equiv U''_k(\rho_0)\) and obtain   \begin{eqnarray} \label{eight}
\frac{d}{dt}\bar{\lambda} &=&
\frac{\partial}{\partial{t}}U''_k(\rho_0) +
U'''_k(\rho_0)\bar{\delta} \nonumber\\ &=&
-v_d(N-1)k^{d-4}\bar{\lambda}^2L_{2,0}^d(0)
-v_dk^{d-4}\left(3\bar{\lambda}+2U'''\rho_0\right)^2{L}_{0,2}^d
(2\bar{\lambda}\rho_0)
\nonumber\\  &&-2v_d(N-1)k^{d-2}\bar{\lambda}Z'L_{2,0}^{d+2}(0)
-2v_dk^{d-2}\left(3\bar{\lambda}+2U'''\rho_0\right)\tilde{Z}'
{L}_{0,2}^{d+2}(2\bar{\lambda}\rho_0)\nonumber\\
&&-v_d(N-1)k^dZ'^2L_{2,0}^{d+4}(0)
-v_dk^d\tilde{Z}'^2{L}_{0,2}^{d+4}(2\bar{\lambda}\rho_0)\nonumber\\
&& +v_{d}(N-1)k^{d-2}U'''L_{1,0}^{d}(0)
+v_dk^{d-2}\left(5U'''+2U^{(4)}\rho_0\right){L}_{0,1}^d
(2\bar{\lambda}\rho_0)
\nonumber\\ &&+v_d(N-1)k^dZ''L_{1,0}^{d+2}(0) +
v_dk^d\tilde{Z}''{L}_{0,1}^{d+2}(2\bar{\lambda}\rho_0) \nonumber\\
&&+ U'''\bar{\delta}. \end{eqnarray} This procedure can be extended
straight forward to higher couplings. Now it is useful to set
\({Z}_k = Z_k(\rho_0,0) \equiv Z_k\) in eq. (\ref{two}) in order to
obtain an effective inverse propagator (for massless fields)
\begin{equation} \label{propagator} Z_k(\rho_0(k),0)q^2 + R_k(q) =
\frac{Z_{k}q^2}{1-f_k^2(q)}. \end{equation} To separate the running
due to dimensions and wave function renormalization from the
running due to interaction we switch to the dimensionless
renormalized couplings (\(Z\equiv Z_k\)) \begin{eqnarray}
\label{dimren} \kappa = k^{2-d}Z\rho_0 &,&\lambda
=k^{d-4}{Z}^{-2}\bar{\lambda},             \nonumber \\   u_3 =
k^{2d-6}{Z}^{-3}U''' &,&u_4 = k^{3d-8}{Z}^{-4}U^{(4)},
\nonumber \\   z_1 = k^{d-2}{Z}^{-2}Z' &,&z_2 =
k^{2d-4}{Z}^{-3}Z'',                \nonumber \\ \tilde{z}_1 =
k^{d-2}{Z}^{-2}\tilde{Z}'&,& \tilde{z}_2 =
k^{2d-4}{Z}^{-3}\tilde{Z}''  \end{eqnarray} and obtain the
\(\beta\)-functions \begin{eqnarray} \label{betakappa} \beta_\kappa
\equiv \frac{\partial{\kappa}}{\partial{t}} &=& (2-d-\eta)\kappa +
2v_d(N-1)(l_1^d + l_1^{d+2}{z_1}/{\lambda}) \nonumber\\
&&+2v_d(3+2u_{3}\kappa/\lambda)l_1^{d}s_1^{d}(2\lambda\kappa) +
2v_{d}l_1^{d+2}s_1^{d+2}(2\lambda\kappa)\tilde{z}_1/\lambda
\end{eqnarray} and \begin{eqnarray} \label{betalambda}
\beta_\lambda \equiv \frac{\partial{\lambda}}{\partial{t}} &=&
(d-4+2\eta)\lambda  +2v_d(N-1)\lambda^2l_2^d
+2v_d\left(3\lambda+2u_3\kappa\right)^2l_2^ds_2^d(2\lambda\kappa)
\nonumber\\  &&-2v_{d}(N-1)u_3l_1^{d}
-2v_d\left(5u_3+2u_4\kappa\right)l_1^ds_1^d(2\lambda\kappa)
\nonumber\\ &&+4v_d(N-1)\lambda{z}_1l_2^{d+2}
+4v_d\left(3\lambda+2u_3\kappa\right)\tilde{z}_1l_2^{d+2}s_2^{d+2}
(2\lambda\kappa)\nonumber\\
&&-2v_d(N-1)z_2l_1^{d+2}
-2v_d\tilde{z}_2l_1^{d+2}s_1^{d+2}(2\lambda\kappa) \nonumber\\
&&+2v_d(N-1)z_1^2l_2^{d+4}
+2v_d\tilde{z}_1^2l_2^{d+4}s_2^{d+4}(2\lambda\kappa) \nonumber\\
&&+u_3\left((d-2+\eta)\kappa+\frac{\partial{\kappa}}{\partial{t}}
\right).
  \end{eqnarray} Here we have used the expressions \begin{eqnarray}
\label{smallL} L_{n,0}^d(0) &=& -2Z^{-n}l_n^d, \nonumber\\
{L}_{0,n}^d(2\bar{\lambda}\rho_0)/L_{n,0}^d(0) &=&
s_n^{d}(\frac{2\bar{\lambda}\rho_0}{k^2}) = s_n^d(2\lambda\kappa).
\end{eqnarray} The anomalous dimension is defined as
\begin{equation} \label{etadef} \eta = -\frac{d}{dt}\ln{Z_k}.
\end{equation} Equation (\ref{betalambda}) is our starting point
for a computation of \(\beta_\lambda\). So far, no approximation
has been made and eq. (\ref{betalambda}) is exact for arbitrary
values of \(\lambda\) and arbitrary dimension \(d\). In addition to
\(\lambda\) it involves, however, several unknown quantities
\footnote{Remember that \(z_1^2\) stands for an appropriate moment
and is in general not equal to the square of \(z_1\).}, namely,
\(\kappa\), \(u_3\), \(u_4\), \(\eta\), \(z_1\), \(z_2\),
\(\tilde{z}_1\), \(\tilde{z}_2\), \(z_1^2\) and \(\tilde{z}_1^2\),
and the threshold functions \(s_n^d(2\lambda\kappa)\). In order to
make contact with the usual \(\beta\)-function we have to compute
all these quantities as functions of \(\lambda\). We will
concentrate on the massless theory which corresponds to the
(approximate) scaling solution of the evolution equation. This
scaling solution is characterized by fixpoints for all couplings
except for the running of \(\lambda\) for \(d=4\). (This running is
very slow for sufficiently small \(\lambda\). The true fixpoint is
the Gaussian for \(\lambda=0\).) For one of the couplings, which we
may associate with \(\kappa\), this fixpoint is infrared unstable.
This corresponds to the relevant parameter of the theory, i.e. the
scalar mass term. For all other couplings (the irrelevant
couplings) the fixpoint is infrared stable and will be approached
rapidly. Due to the existence of the marginal parameter \(\lambda\)
for \(d=4\) the fixpoints for the various couplings depend on
\(\lambda\). Inserting the \(\lambda\)-dependent fixpoints for
\(\kappa\), \(\eta\) etc.  into eq. (\ref{betalambda}) will then
yield \(\beta_\lambda\) as a function of \(\lambda\). Using the
nonperturbative proposal of the last section we compute \(\kappa\),
\(u_3\) and \(\eta\) from corresponding exact evolution equations -
these couplings plus \(\lambda\) correspond to the set \(\{A\}\).
The remaining parameters \(z_1\), \(z_2\), \(\tilde{z}_1\),
\(\tilde{z}_2\), \(z_1^2\) and \(\tilde{z}_1^2\) constitute the set
\(\{B\}\) and are computed using the renormalization group improved
one loop approximation. The latter can be evaluated as functions of
\(\lambda\) and \(\kappa\) in the approximation of a quartic
potential \(U_k\) with uniform \(Z_k=\tilde{Z}_k\). \\ In the
present paper we want to check our method for four dimensions where
the scalar coupling \(\lambda\) is small for \(k\ll\Lambda\) due to
the trivality of the \(\varphi^4\)-theory. We want to compute the
beta function of the quartic coupling \(\lambda\) to order
\(O({\lambda}^3)\), i.e. we want to know the r.h.s. of eq.
(\ref{betalambda}) to order \(O({\lambda}^3)\). Therefore we have
to know the expansion of the following quantities in powers of
\(\lambda\) to the specified order \begin{eqnarray} \label{order}
\kappa \sim O(\lambda^0) &,& u_3 \sim O({\lambda}^3),  \nonumber\\
z_1 \sim O({\lambda}^2)  &,& z_2 \sim O({\lambda}^3), \nonumber\\
\tilde{z}_1 \sim O({\lambda}^2) &,& \tilde{z}_2 \sim
O({\lambda}^3), \nonumber\\ \eta \sim O({\lambda}^2) &.&
\end{eqnarray} From the connection between exact evolution
equations and one loop expressions we recognize that the orders of
the various couplings wich are specified in eq. (\ref{order}) are
also the lowest orders which appear in their perturbative
expansion. To see this, remember that e.g. in leading order of
$\lambda$ and at one loop \(U'''_k(\rho_0)\) is nothing but the
contribution to the six-point-function at zero external momenta in
the symmetric phase and that \(Z'_k(\rho_0)\) is a contribution to
the \(q^2\)-dependence of the 1-loop four-point-function. Similar
arguments hold for the other quantities. We see that higher
derivatives of the potential (like the $\varphi^8$-coupling $u_4$)
or of the wave function renormalization don't contribute to the
beta function of the quartic coupling to order \(O(\lambda^3)\).
Consequently these quantities can be truncated. From the
combination of evolution equations with one loop expressions we get
insight in the important couplings and an efficient tool for
practical calculations. With this knowledge we have to compute the
r.h.s. of \begin{eqnarray} \label{betalambdaref} \beta_\lambda &=&
2\eta\lambda  +\frac{1}{16\pi^2}(N-1)l_2^4\lambda^2
+\frac{9}{16\pi^2}\lambda^{2}l_2^4s_2^4(2\lambda\kappa) \nonumber\\
&&-\frac{1}{16\pi^2}(N-1)u_3l_1^{4}
-\frac{5}{16\pi^2}u_3l_1^4s_1^4(2\lambda\kappa) \nonumber\\
&&+\frac{1}{8\pi^2}(N-1)\lambda{z}_1l_2^{6}
+\frac{3}{8\pi^2}\lambda\tilde{z}_1l_2^{6}s_2^{6}(2\lambda\kappa)
\nonumber\\
&&-\frac{1}{16\pi^2}(N-1)z_2l_1^{6}
-\frac{1}{16\pi^2}\tilde{z}_2l_1^{6}s_1^{6}(2\lambda\kappa)
\nonumber\\    &&+2u_3\kappa + O(\lambda^4) \end{eqnarray} where
the fixed point values have to be inserted for all quantities
except \(\lambda\). So we have to calculate the \(\beta\)-function
of the coupling \(u_3\) and the one loop expressions for the wave
function renormalization constants \(Z_1\), \(Z_2\),
\(\tilde{Z}_1\) and \(\tilde{Z}_2\). Our aim is to reproduce the
known result of the two loop perturbation theory, namely
\begin{equation} \label{total} \beta_\lambda =
\frac{1}{16\pi^2}(N+8){\lambda}^2 -
\frac{1}{(16\pi^2)^2}(9N+42){\lambda}^3. \end{equation} For small
values of \(\lambda\) several simplifications occur. The most
important one is that all couplings contributing to
\(\beta_\lambda\) (\ref{betalambdaref}) in order \(\sim \lambda^3\)
need only be computed in lowest order in \(\lambda\) and that hence
these corrections are simply additive. This means that for the
computation of a given coupling, say \(u_3\), we can put \(z_1\)
etc. to zero. This additivity is not crucial for a numerical
solution of the nonperturbative evolution equations but it helps
considerabley for the analytical study of the present paper. It
allows us to treat the different "corrections" to \(\beta_\lambda\)
seperately and to compare their relative importance. This will be
done in the next sections. A second simplification uses the fact
that the difference between \(\tilde{P}\) and \(P\) is a correction
of order \(\sim \lambda^2\) (\ref{order}). In the order we are
interested in we can always use (\ref{propagator}) \begin{equation}
\label{propfix}
P(x)=\tilde{P}(x)=\frac{Z_k{x}}{1-e^{-\frac{x}{k^2}}}
\end{equation} in the integrals (\ref{five}) and omit the
\(k\)-dependence of \(Z_k\) in \(\frac{\partial{P}}{\partial{t}}\).
This permits a simple analytical computation of the constants
\(l_n^d\) \cite{4}. In particular, one finds \begin{equation}
l_1^4=1, \qquad l_2^4=1, \qquad l_3^4=3\ln\left(\frac{4}{3}\right),
\qquad l_1^6=2, \qquad l_2^6=\frac{3}{2}. \end{equation} Finally a
third simplification concerns the threshold functions
\(s_n^d(2\lambda\kappa)\). They can be expanded for small
\(\lambda\kappa\) according to \begin{equation}
l_n^ds_n^d(2\lambda\kappa) = l_n^d - 2n\lambda\kappa{l}_{n+1}^d +
\dots  \end{equation} We observe that we can put \(s_n^d=1\) except
for the term \(\sim\lambda^2s_n^4\) in eq. (\ref{betalambdaref}).
\setcounter{equation}{0} \section{The average potential with
uniform wave function renormalization} We first consider in this
section a uniform wave function renormalization
\(Z_k(\rho,q^2)=\tilde{Z}_k(\rho,q^2)=Z_k\) and therefore set
\begin{eqnarray} \label{fi.zero} Z'=Z''=\cdots=0, \nonumber\\
Y=Y'=\cdots=0.  \end{eqnarray} In this approximation one obtains
from differentiating eq. (\ref{four}) with respect to \(\rho\)
\begin{eqnarray} \label{fi.one}
\frac{\partial}{\partial{t}}U'''(\rho_0) =
2v_d(N-1)k^{d-6}\bar{\lambda}^3L_{3,0}^d(0) +
54v_dk^{d-6}\bar{\lambda}^3{L}_{0,3}^d(2\bar{\lambda}\rho_0).
\end{eqnarray} The lowest order contribution for the running of
\(\kappa\) and \(u_3\) therefore read  \begin{eqnarray}
\label{fi.two} \frac{\partial{\kappa}}{\partial{t}} &=& (2-d)\kappa
+ 2v_d(N+2)l_1^d, \nonumber\\ \label{fi.twob}
\frac{\partial}{\partial{t}}u_3 &=& (2d-6)u_3 -
4v_d(N+26)l_3^d\lambda^3.  \end{eqnarray} The fixed points are now
easily calculated as the roots of eq. (\ref{fi.two}). One finds in
leading order  \begin{equation} \label{kappafix} \kappa_\star =
2v_d\frac{N+2}{d-2}l_1^d  \end{equation} which is infrared unstable
and   \begin{equation} \label{u3fix} (u_3)_\star =
2v_d\frac{N+26}{d-3}l_3^d\lambda^3. \end{equation} Note that the
last value coincidences with the one loop expression for \(u_3\)
\begin{equation} \label{u3loop}
u_3^{(1)}=2(N+26)v_d\lambda^3\int\limits_0^\infty{dx}x^{\frac{d}{2}
-1}
P^{-3}(x)
\end{equation} which follows from eq. (\ref{split}) at zero field
\(\varphi=0\) and \(n=6\). This coincidence can be understood by
looking at the estimate of error (\ref{error}) of the
\(\beta\)-function  caused by the use of "renormalization group
improved" one loop expression (\ref{u3loop}) instead of the
evolution equation (\ref{fi.twob}). Comparing the estimate of error
for \(u_3\)  \begin{equation}
\frac{\partial{\Delta{u_3}}}{\partial{t}}=
\frac{\partial}{\partial{t}}
\Delta\Gamma_k^{(6)}[\varphi=0]
=
-\frac{1}{2}{T}{r}\frac{\delta^6}{\delta\varphi_1(q=0)\cdots\delta
\varphi_6(q=0)}\left\{\left(\Gamma_{k}^{(2)}+{R}_{k}\right)^{-1}
\frac{\partial}{\partial{t}}{\Gamma}_{k}^{(2)}\right\}
\Bigg\vert_{\varphi=0}
\end{equation} with its exact evolution equation \begin{equation}
\frac{\partial{u_3}}{\partial{t}}=\frac{\partial}{\partial{t}}
\Gamma_{k}^{(6)}[\varphi=0]=
\frac{1}{2}{T}{r}\frac{\delta^6}{\delta\varphi_1(q=0)\cdots\delta
\varphi_6(q=0)}\left\{(\Gamma_{k}^{(2)}[\varphi]+{R}_{k})^{-1}
\frac{\partial}{\partial{t}}
{R}_{k}\right\}\Bigg\vert_{\varphi=0} \end{equation} we recognize
that the estimated error is of higher order in \(\lambda\) since
\(\frac{\partial}{\partial{t}}\Gamma_k^{(2)}[0]\sim\lambda^2\)
whereas \(\frac{\partial}{\partial{t}}R_k\sim\lambda^0\). We also
observe that the difference between \(\Gamma_k^{(6)}[\varphi=0]\)
and \(\Gamma_k^{(6)}\) evaluated at the minimum is of higher order
in \(\lambda\). It is therefore no surprise that the lowest order
contributions eqs. (\ref{u3fix}), (\ref{u3loop}) coincide.
Inserting the fixed points (\ref{kappafix}), (\ref{u3fix}) in eq.
(\ref{betalambdaref}) we obtain in four dimensions \begin{equation}
\beta_\lambda = \frac{1}{16\pi^2}(N+8)\lambda^2 -
\frac{1}{(16\pi^2)^2}l_1^4l_3^4\left(20N+88\right)\lambda^3
\end{equation} which is numerically \begin{equation} \beta_\lambda
= \frac{1}{16\pi^2}(N+8)\lambda^2 -
\frac{1}{(16\pi^2)^2}\left(17.26N+75.95\right)\lambda^3
\end{equation} to be compared with eq. (\ref{total}).
\setcounter{equation}{0} \section{Field and momentum dependence of
the wave function renormalization} We next take the field and
momentum dependence of the wave function renormalization into
account. We want to compute \(z_1\), \(\tilde{z}_1\), \(z_2\) and
\(\tilde{z}_2\) in a renormalization group improved one loop
approximation. For this purpose we write \(Z_k(\rho,Q^2)-Z_k(0,0)\)
as an appropriate difference of two point functions and similar for
\(\tilde{Z}_k\). This difference is evaluated according to
(\ref{split}) where on the r.h.s. an approximation for
\(\Gamma_k^{(2)}\) is infered from the ansatz (\ref{expan}) with
\(Z_k(\rho,-\partial^2)=Z_k\) and \(Y_k(\rho,-\partial^2)=0\). In a
first approach we neglect the momentum dependence and approximate
\(Z_k(\rho)\) by \(Z_k(\rho,q^2=0)\). The one loop expressions are
derived in appendix A and read \begin{eqnarray} \label{olappzeta}
Z_k(\rho,0) &=& {Z}_k(0,0)
+8\frac{v_d}{d}k^{d-6}\left(U''_k(\rho)\right)^2\rho{E}_{2,2}^d
(w_1,w_2),
\nonumber\\ \tilde{Z}_k(\rho,0) &=& {{Z}_k}(0,0) +
4\frac{v_d}{d}k^{d-6}\left(3U''_k(\rho)+2U'''_k(\rho)\rho\right)^2
\rho{E}_{0,4}^d(w_1,w_2)
\nonumber\\
&&+4(N-1)\frac{v_d}{d}k^{d-6}\left(U''_k(\rho)\right)^2
\rho{E}_{4,0}^d(w_1,w_2).
\end{eqnarray} Here we have introduced the dimensionless integrals
\begin{equation} E_{n_1,n_2}^d(w_1,w_2) =
k^{2(n_1+n_2-1)-d}\int_{0}^\infty{d}xx^{\frac{d}{2}}\dot{P}^2
(P+w_1)^{-n_1}(P+w_2)^{-n_2}
\end{equation} where \(w_{1,2}\) are given by (\ref{double_u}) and
we have used \(x=q^2\), \(\dot{P}=\frac{\partial{P}}{\partial{x}}\)
with \(P\) given by (\ref{propfix}). The \(\rho\)-derivatives of
eqs. (\ref{olappzeta}) at fixed \(\rho_0\) yield in lowest order
\begin{eqnarray} Z'&=& 8\frac{v_d}{d}k^{d-6}\bar{\lambda}^2{E}_4^d,
 \nonumber\\ Z''&=& -128\frac{v_d}{d}k^{d-8}\bar{\lambda}^3{E}_5^d,
\nonumber\\ \tilde{Z}'&=&
4(N+8)\frac{v_d}{d}k^{d-6}\bar{\lambda}^2{E}_4^d, \nonumber\\
\tilde{Z}''&=& -32(N+26)\frac{v_d}{d}k^{d-8}\bar{\lambda}^3{E}_5^d
\end{eqnarray} with \(E_{n_1+n_2}^d \equiv E_{n_1,n_2}^d(0,0)\). We
switch to the dimensionless renormalized quantities and find for
\(d=4\) \begin{eqnarray} \label{olex} z_1 &=&
\frac{1}{16\pi^2}\lambda^2{\epsilon}_4^4, \nonumber\\ z_2 &=&
-\frac{1}{\pi^2}\lambda^3{\epsilon}_5^4, \nonumber\\ \tilde{z}_1
&=& \frac{N+8}{32\pi^2}\lambda^2{\epsilon}_4^4, \nonumber\\
\tilde{z}_2 &=& -\frac{N+26}{4\pi^2}\lambda^3{\epsilon}_5^4
\end{eqnarray} where \begin{equation} E_n^d =
Z^{2-n}\epsilon_n^d\quad \mbox{and}\quad
\epsilon_4^4=\frac{1}{2},\qquad
\epsilon_5^4=\frac{1}{2}\ln\left(\frac{4}{3}\right). \end{equation}
Again these one loop expressions are in lowest order equal to the
appropriate fixpoints of the corresponding evolution equations. As
before this holds because we include only the leading order
contributions to the one loop expressions or the appropriate
evolution equations.\\ The formulae (\ref{olex}) are only
approximations and do not include the full contributions in leading
order in \(\lambda\). For a precision estimate in this order we
have to use approximately weighted functions as indicated by eqs.
(\ref{moma}), (\ref{momb}). We therefore have to take the momentum
dependence of \(Z_k(\rho,q^2)\) into account and employ the
renormalization group improved one loop expressions (\ref{loopq})
given in appendix A. We notice that
\(\frac{Z_k(\rho_0,Q^2)}{Z_k}=1+O(\bar{\lambda}^2)\). This is the
well known result from standard perturbation theory. In the
\(\beta_\lambda\)-function (\ref{betalambdaref}) the quantities
\(Z',\tilde{Z}'\) enter up to order \(O(\lambda^2)\) and
\(Z'',\tilde{Z}''\) enter up to order \(O(\lambda^3)\). Deriving
equations (\ref{loopq}) with respect to \(\rho\) at the minimum
\(\rho_0\) we obtain in leading order \begin{eqnarray}
Z'_k(\rho_0,Q^2) &=&
-2\frac{\bar{\lambda}^2}{Q^2}(2\pi)^{-d}\int{d}^{d}{q}P^{-1}(q)
\left[P^{-1}(q+Q)-P^{-1}(q)\right],\nonumber\\
Z''_k(\rho_0,Q^2) &=&
16\frac{\bar{\lambda}^3}{Q^2}(2\pi)^{-d}\int{d}^{d}{q}P^{-2}(q)
\left[P^{-1}(q+Q)-P^{-1}(q)\right],\nonumber\\
\tilde{Z}'_k(\rho_0,Q^2) &=&
-(N+8)\frac{\bar{\lambda}^2}{Q^2}(2\pi)^{-d}\int{d}^{d}{q}P^{-1}(q)
\left[P^{-1}(q+Q)-P^{-1}(q)\right],
\nonumber\\ \tilde{Z}''_k(\rho_0,Q^2) &=&
4(N+26)\frac{\bar{\lambda}^3}{Q^2}(2\pi)^{-d}\int{d}^{d}{q}P^{-2}(q)
\left[P^{-1}(q+Q)-P^{-1}(q)\right].
 \end{eqnarray} We notice that in the required order in \(\lambda\)
the mass term \(\sim2\lambda\kappa{k}^2\) in the denominators can
be neglected and we have only to solve the one loop integrals for a
massless theory. We evaluate these integrals using the Laplace
transform \begin{equation} \label{Schwinger} P^{-1}(q) =
\frac{1-e^{-\frac{q^2}{k^2}}}{q^2} =
\int\limits_0^{k^{-2}}d\alpha{e}^{-\alpha{q}^2} \end{equation} and
obtain \begin{eqnarray} \label{Iinteg} I(Q^2) &\equiv &
\frac{1}{2}(2\pi)^{-d}\int{d}^{d}{q}P^{-1}(q)\left[P^{-1}(q+Q)-
P^{-1}(q)\right]
\nonumber\\ &=&
v_{d}\Gamma\left(\frac{d}{2}\right)\int\limits_0^{k^{-2}}
\int\limits_0^{k^{-2}}d\alpha{d}\beta\left(e^{-Q^2
\frac{\alpha\beta}{\alpha+\beta}}-1\right)
(\alpha+\beta)^{-\frac{d}{2}}
\end{eqnarray} and \begin{eqnarray} \label{Jinteg} J(Q^2) &\equiv &
\frac{1}{2}(2\pi)^{-d}\int{d}^{d}{q}P^{-2}(q)\left[P^{-1}(q+Q)-
P^{-1}(q)\right]
\nonumber\\ &=&
v_{d}\Gamma\left(\frac{d}{2}\right)\int\limits_0^{k^{-2}}
\int\limits_0^{k^{-2}}\int\limits_0^{k^{-2}}d\alpha{d}\beta{d}
\gamma\left(e^{-Q^2\frac{\alpha(\beta+\gamma)}{\alpha+\beta+
\gamma}}-1\right)(\alpha+\beta+\gamma)^{-\frac{d}{2}}.
\end{eqnarray} After some transformations we obtain in four
dimensions (See app. B) \begin{equation} \label{Iintegral} I(Q^2) =
v_{4}\sum_{n=0}^\infty\frac{(-1)^{n+1}}{(n+2)!(n+1)}
\left(\frac{1}{2}\right)^{n+1}\left(\frac{Q^2}{k^2}\right)^{n+1}
\end{equation} and \begin{equation} \label{Jintegral} J(Q^2) =
v_{4}k^{-2}\sum_{n=0}^\infty\frac{(-1)^n}{(n+2)!(n+1)}
\left(\frac{Q^2}{k^2}\right)^{n+1}\left\{\int_0^\frac{1}{2}dz
\frac{z^{n+1}}{(1-z)^2}
-
\int_{\frac{1}{2}}^{\frac{2}{3}}dz\frac{z^{n+1}}{(1-z)^2}\right\}.
\end{equation} The dimensionless renormalized quantities read in
lowest order in \(\lambda\) \begin{eqnarray} z_1(Q^2) &=&
2v_{4}\lambda^2\sum_{n=0}^\infty\frac{(-1)^n}{(n+2)!(n+1)}
\left(\frac{1}{2}\right)^{n}\left(\frac{Q^2}{k^2}\right)^{n},
 \nonumber\\ \tilde{z}_1(Q^2) &=&
(N+8)v_{4}\lambda^2\sum_{n=0}^\infty\frac{(-1)^n}{(n+2)!(n+1)}
\left(\frac{1}{2}\right)^{n}\left(\frac{Q^2}{k^2}\right)^{n},
\nonumber\\ z_2(Q^2) &=&
32v_{4}\lambda^3\sum_{n=0}^\infty\frac{(-1)^n}{(n+2)!(n+1)}
\left(\frac{Q^2}{k^2}\right)^{n}\left\{\int\limits_0^\frac{1}{2}dz
\frac{z^{n+1}}{(1-z)^2}
-
\int\limits_{\frac{1}{2}}^{\frac{2}{3}}dz
\frac{z^{n+1}}{(1-z)^2}\right\},
 \nonumber\\ \tilde{z}_2(Q^2) &=&
8(N+26)v_{4}\lambda^3\sum_{n=0}^\infty\frac{(-1)^n}{(n+2)!(n+1)}
\left(\frac{Q^2}{k^2}\right)^{n}\left\{\int\limits_0^\frac{1}{2}dz
\frac{z^{n+1}}{(1-z)^2}
-
\int\limits_{\frac{1}{2}}^{\frac{2}{3}}dz
\frac{z^{n+1}}{(1-z)^2}\right\}.
\nonumber  \end{eqnarray} Notice that the (\(n=0\))-terms of the
sum correspond to eqs. (\ref{olex}). We use these expressions to
calculate the appropriate moments (\ref{moma}) which occur in the
\(\beta_\lambda\)-function (\ref{betalambdaref}).  \begin{eqnarray}
\label{zmoments} z_1l_2^6 &=&
2v_{4}\lambda^2\sum_{n=0}^\infty\frac{(-1)^n}{(n+2)!(n+1)}
\left(\frac{1}{2}\right)^{n}l_2^{6+2n}\nonumber\\
 &=& (24\ln3+16\ln2-20\ln5-4)v_4\lambda^2,\nonumber\\
\tilde{z}_1l_2^6 &=&
(N+8)v_{4}\lambda^2\sum_{n=0}^\infty\frac{(-1)^n}{(n+2)!(n+1)}
\left(\frac{1}{2}\right)^{n}l_2^{6+2n}\nonumber\\
&=&(N+8)(12\ln3+8\ln2-10\ln5-2)v_4\lambda^2, \nonumber\\ z_2l_1^6
&=&
32v_{4}\lambda^3\sum_{n=0}^\infty\frac{(-1)^n}{(n+2)!(n+1)}l_1^{6+2n}
\left\{\int_0^\frac{1}{2}dz\frac{z^{n+1}}{(1-z)^2}
- \int_{\frac{1}{2}}^{\frac{2}{3}}dz\frac{z^{n+1}}{(1-z)^2}\right\}
\nonumber\\ &=&16(12\ln3-5\ln5-8\ln2)v_4\lambda^3, \nonumber\\
\tilde{z}_2l_1^6 &=&
8(N+26)v_{4}\lambda^3\sum_{n=0}^\infty\frac{(-1)^n}{(n+2)!(n+1)}
l_1^{6+2n}
\left\{\int_0^\frac{1}{2}dz\frac{z^{n+1}}{(1-z)^2}
- \int_{\frac{1}{2}}^{\frac{2}{3}}dz\frac{z^{n+1}}{(1-z)^2}\right\}
\nonumber\\ &=& 4(N+26)(12\ln3-5\ln5-8\ln2)v_4\lambda^3.
\end{eqnarray} (See app. C for the evaluation of the sums.)
Comparing with the expressions (\ref{olex}) we see that the
\(Q^2\)-dependence of the wave function accounts for 21\%, 21\%,
30\% and 30\% of the moments \(z_1\), \(\tilde{z}_1\), \(z_2\) and
\(\tilde{z}_2\), respectively. Inserting the results
(\ref{zmoments}) into the \(\beta_\lambda\)-function
(\ref{betalambdaref}) reads  \begin{equation} \label{buteta}
\beta_\lambda = \frac{N+8}{16\pi^2}\lambda^2 -
\frac{10N+44}{(16\pi^2)^2}\lambda^3 + 2\eta\lambda. \end{equation}
In order to compute the complete coefficient \(\sim\lambda^3\) we
only need now the anomalous dimension \(\eta\) in order
\(\lambda^2\). \setcounter{equation}{0} \section{The anomalous
dimension} The anomalous dimension \(\eta\) is defined by
identifying \(Z_k\) with \(Z_k(\rho_0(k),Q^2=0)\) \begin{eqnarray}
\label{etadef2} \eta &=& -\frac{d}{d{t}}\ln{Z}_k(\rho_0)
\nonumber\\ &=&
-Z^{-1}_k(\rho_0)\frac{\partial{Z_k(\rho_0)}}{\partial{t}}-
Z^{-1}_k(\rho_0)Z'_k(\rho_0)\bar{\delta}
\end{eqnarray} where  \begin{eqnarray} Z_k(\rho) \equiv
Z_k(\rho,Q^2=0) &=&
\Omega^{-1}\lim_{Q^2\to{0}}\frac{\partial}{\partial{Q^2}}
\frac{\delta^2}{\delta\varphi^2(Q)\delta\varphi_2^\star(Q)}
\Gamma_k\bigg\vert_{\rho}
\nonumber\\ &=&
\lim_{Q^2\to{0}}\frac{\partial}{\partial{Q^2}}\Gamma_k^{(2)}
(Q,-Q;\rho)
\end{eqnarray} is evaluated for a constant field configuration.
With the exact evolution equation eq. (\ref{master}) we obtain an
exact equation for the \(k\)-dependence of \(Z_k(\rho)\) (for
\(N\ge 2\) \footnote{For \(N=1\) one should use functional
derivatives with respect to \(\varphi\equiv\varphi_1\) . This
changes the formulae of this section but leads to the same final
result for \(\eta\).}) \begin{equation} \label{fullzeta}
\frac{\partial{Z_k(\rho)}}{\partial{t}} =
\frac{1}{2}\Omega^{-1}\lim_{Q^2\to{0}}
\frac{\partial}{\partial{Q^2}}
\frac{\delta^2}{\delta\varphi^2(Q)\delta\varphi_2^\star(Q)}
Tr\left\{\left[\Gamma_k^{(2)}
+ R_k\right]^{-1}\partial_t{R}_k\right\}\bigg\vert_{\rho}.
\end{equation} We perform the functional derivatives (omitting the
internal field indices) \begin{eqnarray} \label{exzet}
\frac{\partial{Z_k(\rho)}}{\partial{t}} &=&
\frac{1}{2}\Omega^{-1}\lim_{Q^2\to{0}}\frac{\partial}{\partial{Q^2}}
\frac{\delta^2}{\delta\varphi^2(Q)\delta\varphi_2^\star(Q)}Tr
\left\{\left[\Gamma_k^{(2)}(q,-q)+R_k(q)\right]^{-1}\partial_{t}
{R}_k(q)\right\}\bigg\vert_{\rho}
\nonumber\\
&=&\frac{1}{2}\Omega^{-1}\lim_{Q^2\to{0}}
\frac{\partial}{\partial{Q^2}}
Tr\bigg\{-\left[\Gamma_k^{(2)}(q,-q)+{R}_k(q)\right]^{-1}\nonumber\\
&&\cdot\Gamma_k^{(4)}(Q,-Q,q,-q)\left[\Gamma_k^{(2)}(q,-q)+
{R}_k(q)\right]^{-1}\partial_{t}{R}_k(q)
\nonumber\\
&&+2\left[\Gamma_k^{(2)}(q,-q)+{R}_k(q)\right]^{-1}
\Gamma_k^{(3)}(Q,q,-q-Q)\left[\Gamma_k^{(2)}(q+Q,-q-Q)
+{R}_k(q+Q)\right]^{-1}\nonumber\\
&&\cdot\Gamma_k^{(3)}(-Q,q+Q,-q)\left[\Gamma_k^{(2)}(q,-q)
+{R}_k(q)\right]^{-1}\partial_{t}{R}_k(q)\bigg\}\bigg\vert_{\rho}
\end{eqnarray} with \begin{equation} \Gamma_k^{(n)}(q_1,\ldots,q_n)
=
\Omega^{-1}\frac{\delta^n}{\delta\varphi(q_1)\cdots\delta
\varphi(q_n)}
\Gamma_k
\end{equation} containing a factor \(\sim \delta(q_1+\dots+q_n)\).
Inserting \(\rho=\rho_0\) in eq. (\ref{exzet}) and combining eq.
(\ref{etadef2}) with eq. (\ref{exzet}) gives an exact
nonperturbative expression for the anomalous dimension \(\eta\) !
It involves the \(k\)-dependent full propagator
\(\left(\Gamma_k^{(2)}+R_k\right)^{-1}\), the 1PI three point
function for arbitrary momenta and the 1PI four point function for
two pairs of opposite momenta. One of the momenta (corresponding to
\(Q^2\)) can be taken small and only an expression quadratic in
this momentum is needed. (We observe that eq. (\ref{exzet}) can
easily be generalized into an exact evolution equation for the full
momentum dependent "wave function renormalization"
\(Z_k(\rho,Q^2)\) by omitting the operation
\(\lim_{Q^2\to{0}}\frac{\partial}{\partial{Q^2}}\), subtracting
from the now \(Q\) dependent r.h.s. the value for \(Q=0\) and
deviding by \(Q^2\), i.e. by combining eq. (\ref{master}) and eq.
(\ref{9.2}).)\\ The ansatz (\ref{expan}) only describes
\(\Gamma_k^{(3)}(0,q,-q)\) and \(\Gamma_k^{(4)}(0,0,q,-q)\) or
other combinations with only two nonvanishing momenta since higher
derivative terms of the structure
\(\partial^2\varphi\partial\varphi\partial\varphi\) and
\((\partial\varphi)^4\) are not included. In a first step we
neglect the momentum dependence of \(\Gamma_k^{(3)}\) and
\(\Gamma_k^{(4)}\) except for the contributions quadratic in the
momenta. We can then determine the r.h.s. of eq. (\ref{exzet}) from
the ansatz (\ref{expan}) with \(Z_k\) and \(Y_k\) depending only on
\(\rho\). In this limit \(\eta\) has been computed in
\cite{4},\cite{13} and we quote only the results. The scale
dependence of \(Z_k(\rho)\) at fixed \(\rho\) reads in this
approximation \begin{equation} \label{zeta}
\frac{\partial{Z_k(\rho)}}{\partial{t}}\bigg\vert_\rho =
\frac{\partial{Z_k(\rho)^{(a)}}}{\partial{t}}\bigg\vert_\rho +
\frac{\partial{Z_k(\rho)^{(b)}}}{\partial{t}}\bigg\vert_\rho
\end{equation} with \begin{eqnarray} \label{azeta}
\frac{\partial{Z_k(\rho)^{(a)}}}{\partial{t}}\bigg\vert_\rho &=&
v_d\left[(N-1)Z'_k(\rho)+Y_k(\rho)\right]k^{d-2}L_{1,0}^d(w_1)
\nonumber\\
 &&+
v_d\left[Z'_k(\rho)+2Z''_k(\rho)\rho\right]k^{d-2}{L}_{0,1}^d(w_2),
 \\ \label{bzeta}
\frac{\partial{Z_k(\rho)^{(b)}}}{\partial{t}}\bigg\vert_\rho &=&
4v_d\left(U''_k(\rho)\right)^2\rho{k}^{d-6}Q_{2,1}^{d,0}(w_1,w_2) +
4v_dY_k(\rho)U''_k(\rho)\rho{k}^{d-4}Q_{2,1}^{d,1}(w_1,w_2)
\nonumber \\ &&+
v_d\left(Y_k(\rho)\right)^2\rho{k}^{d-2}Q_{2,1}^{d,2}(w_1,w_2) -
8v_dZ'_k(\rho)U''_k(\rho)\rho{k}^{d-4}L_{1,1}^{d}(w_1,w_2)
\nonumber \\
&&-\frac{4v_d}{d}\left(Z'_k(\rho)\right)^2\rho{k}^{d-2}
L_{1,1}^{d+2}(w_1,w_2)
- 4v_dZ'_k(\rho)Y_k(\rho)\rho{k}^{d-2}L_{1,1}^{d+2}(w_1,w_2)
\nonumber \\
&&+\frac{16v_d}{d}Z'_k(\rho)U''_k(\rho)\rho{k}^{d-4}
N_{2,1}^{d}(w_1,w_2)\nonumber\\
 &&+
\frac{8v_d}{d}Z'_k(\rho)Y_k(\rho)\rho{k}^{d-2}N_{2,1}^{d+2}(w_1,w_2).
 \end{eqnarray} Note that eq. (\ref{azeta}) stems from the part of
(\ref{exzet}) which involves the four point function whereas eq.
(\ref{bzeta}) is originated from the part of (\ref{exzet}) which
involves the three point functions. The integrals \(N_{n_1,n_2}^d\)
are defined by, with \(\dot{P}=\frac{\partial{P}}{\partial{x}}\)
\begin{equation} N_{n_1,n_2}^d(w_1,w_2) =
k^{2(n_1+n_2-1)-d}\int_0^{\infty}dxx^{d\over{2}}
\frac{\partial}{\partial{t}}\left\{\dot{P}(P+w_1)^{-n_1}
(\tilde{P}+w_2)^{-n_2}\right\}
\end{equation}  and the integrals \begin{eqnarray}
Q_{n_1,n_2}^{d,\alpha}(w_1,w_2) &=&
k^{2(n_1+n_2-\alpha)-d}\int_0^{\infty}dxx^{\frac{d}{2}-1+\alpha}
\nonumber\\
&&\cdot\frac{\partial}{\partial{t}}\left\{\left[\dot{P}+
\frac{2x}{d}\ddot{P}-\frac{4x}{d}\dot{P}^{2}(P+w_1)^{-1}\right]
(P+w_1)^{-n_1}(\tilde{P}+w_2)^{-n_2}\right\}
\end{eqnarray} are related by partial integration to other
integrals \begin{equation} \label{mint} M_{n_1,n_2}^d(w_1,w_2) =
k^{2(n_1+n_2-1)-d}\int_0^{\infty}dxx^{d\over{2}}
\frac{\partial}{\partial{t}}\left\{\dot{P}^{2}(P+w_1)^{-n_1}
(\tilde{P}+w_2)^{-n_2}\right\}
\end{equation} through \begin{eqnarray} \label{qmrel}
Q_{n_1,n_2}^{d,\alpha}(w_1,w_2) &=&
\frac{2n_1-4}{d}M_{n_1+1,n_2}^{d+2\alpha}(w_1,w_2) +
\frac{2n_2}{d}M_{n_1,n_2+1}^{d+2\alpha}(w_1,w_2) \nonumber \\
&&+\frac{2n_2}{d}\rho{Y}_k(\rho)N_{n_1,n_2+1}^{d+2\alpha}(w_1,w_2)
- \frac{2\alpha}{d}N_{n_1,n_2}^{d+2\alpha-2}(w_1,w_2).
\end{eqnarray} Note, that the \(E_{n_1,n_2}^d\)-integrals are
closely related to the \(M_{n_1,n_2}^d\)-integrals since
\(\frac{\partial}{\partial{t}}g(\frac{x}{k^2}) =
-2x\frac{\partial}{\partial{x}}g(\frac{x}{k^2})\) for an arbitrary
function depending only on the ratio \(\frac{x}{k^2}\). In lowest
order of \(\lambda\) eq. (\ref{zeta}) reads at the minimum
\(\rho_0\)   \begin{equation}
\frac{\partial{Z}}{\partial{t}}=v_d(NZ'+Y)k^{d-2}L^d_{1,0}(0)+
\frac{8}{d}v_d\bar{\lambda}^2\rho_0k^{d-6}M^d_{4,0}(0).
\end{equation} Expressed in the dimensionless renormalized
quantities (and \(y=Z^{-2}k^{d-2}Y\)) one obtaines for the
anomalous dimension (\ref{etadef2}) \begin{equation}
\eta=16\frac{v_d}{d}m_4^d\kappa\lambda^2+2v_d(Nz_1+y)l_1^d-
z_1\left(\frac{\partial{\kappa}}{\partial{t}}+(d-2)\kappa\right)
\end{equation} where \begin{equation} \label{smallM} m_n^d =
-\frac{1}{2}Z^{n-2}M^d_{n,0}(0). \end{equation} We remember
\(y=\tilde{z}_1-z_1\) and write at the fixpoint of our theory
\begin{equation} \label{etaold} \eta =
16\frac{v_d}{d}m_4^d\kappa_\star\lambda^2+
2v_d{l}_1^d(\tilde{z}_1)_\star-6v_d{l}_1^{d}(z_1)_\star.
\end{equation} Since we have neglected the momentum dependence of
\(Z_k(\rho,Q^2)\) the appropriate one loop expressions are given by
(\ref{olex}) and we obtain in the present approximation and for
\(d=4\) \begin{eqnarray} \label{missdelta} \eta =
\frac{3(N+2)}{(32\pi^2)^2}\lambda^2. \end{eqnarray} Next we include
in eq. (\ref{exzet}) the momentum dependence of \(\Gamma_k^{(3)}\)
and \(\Gamma_k^{(4)}\) beyond the approximation quadratic in
momenta. We observe that the momentum dependence of the three point
and four point function is always of higher order in \(\lambda\)
compared with the momentum independent part. The contribution
\(\sim \left(\Gamma_k^{(3)}\right)^2\) is already \(\sim\lambda^2\)
and we therefore do not need the momentum dependence of the three
point function for a computation of \(\eta\) in order
\(\lambda^2\). On the other hand we need the quantity
\begin{eqnarray} \label{rest} \Delta_k(Q,-Q,q,-q,) &=&
\Gamma_k^{(4)}(Q,-Q,q,-q)-\Gamma_k^{(4)}(0,0,q,-q)\nonumber\\
&&-\Gamma_k^{(4)}(Q,-Q,0,0)+\Gamma_k^{(4)}(0,0,0,0) \end{eqnarray}
in order \(\lambda^2\). This summarizes the neglected terms and the
insertion of (\ref{rest}) into (\ref{exzet}) gives the complete
correction to \(\eta\) in order \(\lambda^2\). In fact, the
relevant \(Q\)-dependence of the first term on the r.h.s. of eq.
(\ref{exzet}) is given by an expansion of \(\Gamma_k^{(4)}\)
quadratic in \(Q\). The most general form of the \(Q^2\) part of
\(\Gamma_k^{(4)}(Q,-Q,q,-q)\) can be written as \begin{equation}
\Gamma_k^{(4)}(Q,-Q,q,-q)=F+G(q^2+Q^2)+H_0(q^2)+H_1(q^2)Q^2+
H_2(q^2)(qQ)^2+O(Q^4)
\end{equation} with  \begin{eqnarray} H_0(0)=0 ,\qquad
\dot{H}(0)=0,\qquad H_1(0)=0. \end{eqnarray} The first two parts
involving \(F\) and \(G\) have already been included in the
approximation used in the first half of this section whereas
\(H_0(q^2)\) does not contribute to \(\eta\). It is easy to verify
that the remaining term \(H_1(q^2)Q^2+H_2(q^2)(qQ)^2\) is just the
part quadratic in an expansion of \(\Delta_k(Q,-Q,q,-q)\). As an
illustration of this argument we give the expansion of
\(\Gamma_k^{(4)}\) in fourth order in the momenta for \(N=1\)
\begin{eqnarray} \Gamma_k^{(4)}(Q,-Q,q,-q) &=&
3U''_k(\rho)+2\rho{U}'''_k(\rho)+4\rho^2{U}^{(4)}_k(\rho)
\nonumber\\ &&+(q^2+Q^2)(Z'_k(\rho)+2\rho{Z}''_k(\rho))\nonumber\\
&&+(q^{2}Q^2+2(qQ)^2)W_k(\rho) \end{eqnarray} and note that the
last term \(\sim{W}_k(\rho)\) corresponds to the contribution from
\(\Delta_k\) neglected so far.\\ We will consider \(\Delta_k\) as
one of the "secondary" couplings in the sense of sect. 2 and
evaluate it by a "renormalization group improved" one loop
expression. In lowest order \(\lambda^2\) we can neglect the mass
terms \(\sim 2\lambda\kappa{k}^2\) in the propagators and only need
a one loop calculation for the massless case, assuming a purely
quartic potential and a constant wave function renormalization
\(Z_k\). With these approximations the one loop expression for the
four point function reads \begin{eqnarray} \label{fourpoint}
\left(\Gamma_k^{(4)}(Q,-Q,q,-q)\right)_{22aa} &=&
-\bar{\lambda}^{2}diag\Big((N+4)A+2B(q,Q), (N+8)A+(N+8)B(q,Q),
\nonumber\\
&&\underbrace{(N+4)A+2B(q,Q),\ldots,(N+4)A+2B(q,Q)}_{N-2}\Big)
\end{eqnarray} where \begin{eqnarray} A &=&
\frac{1}{2}(2\pi)^{-d}\int{d}^{d}{p}P^{-2}(p),\nonumber\\ B(q,Q)
&=&
\frac{1}{2}(2\pi)^{-d}\int{d}^{d}{p}P^{-1}(p)\left[P^{-1}(p-Q-q)+
P^{-1}(p-Q+q)\right].
\end{eqnarray} The expressions \(A\) and \(B\) are both ultraviolet
divergent in four dimensions but these divergencies cancel for the
difference (\ref{rest}) \begin{eqnarray} \Delta_k(Q,-Q,q,-q,) =
-\bar{\lambda}^2{diag}\left(2,N+8,\overbrace{2,\ldots,2}^{N-2}\right)
\frac{1}{2}(2\pi)^{-d}\int{d}^{d}{p}P^{-1}(p)\nonumber\\
\cdot\left[2P^{-1}(p)+P^{-1}(p-Q-q)+P^{-1}(p-Q+q)-2P^{-1}(p+Q)-
2P^{-1}(p+q)\right]
\end{eqnarray} which is finite in any dimension less than six.
Again we evaluate this expression using the Laplace transform
(\ref{Schwinger}) and obtain \begin{eqnarray} \Delta_k(Q,-Q,q,-q,)
=
-v_d\Gamma\left(\frac{d}{2}\right)\bar{\lambda}^{2}diag\left(2,N+8,
\overbrace{2,\ldots,2}^{N-2}\right)\int\limits_0^{k^{-2}}
\int\limits_0^{k^{-2}}d\alpha{d}\beta\nonumber\\
\cdot\bigg[e^{-\frac{\alpha\beta}{\alpha+\beta}(Q+q)^2}-1+
e^{-\frac{\alpha\beta}{\alpha+\beta}(Q-q)^2}-1-
2e^{-\frac{\alpha\beta}{\alpha+\beta}Q^2}+2-
2e^{-\frac{\alpha\beta}{\alpha+\beta}Q^2}+2\bigg]
(\alpha+\beta)^{-\frac{d}{2}}.
\end{eqnarray}  With the \(I(Q^2)\)-integral (\ref{Iintegral}) we
write in four dimensions \begin{eqnarray} \Delta_k(Q,-Q,q,-q,) =
-v_4\bar{\lambda}^{2}diag\left(2,N+8,
\overbrace{2,\ldots,2}^{N-2}\right)
\sum_{n=0}^\infty\frac{(-1)^{n+1}}{(n+2)!(n+1)}
\left(\frac{1}{2}\right)^{n+1}\nonumber\\
\cdot\Bigg[\left(\frac{(Q+q)^2}{k^2}\right)^{n+1}+
\left(\frac{(Q-q)^2}{k^2}\right)^{n+1}-
2\left(\frac{Q^2}{k^2}\right)^{n+1}-
2\left(\frac{q^2}{k^2}\right)^{n+1}\Bigg].
\end{eqnarray} We now insert this expression into eq.
(\ref{fullzeta}), perform the
\(\lim_{Q^2\to{0}}\frac{\partial}{\partial{Q^2}}\) operation under
the momentum integration corresponding to the trace and obtain with
(\ref{etadef2}) the contribution \begin{eqnarray} \label{deltaeta}
\Delta\eta &=&
-3(N+2)v_4^2\lambda^2\int_0^\infty{d}y\partial_{t}P(y)P^{-2}(y)
\sum_{n=1}^\infty\frac{(-1)^{n+1}}{(n+1)!}
\left(\frac{y}{2}\right)^{n+1}
\nonumber\\ &=& -\frac{N+2}{(32\pi^2)^2}\lambda^2 \end{eqnarray}
where we set \(y=q^2/k^2\) and used eq. (\ref{propfix}) to perform
the integration. Adding this contribution \(\Delta\eta\) to the
result from eq. (\ref{missdelta}) we obtain in order \(\lambda^2\)
\begin{equation} \label{2leta}
\eta=\frac{N+2}{2}\frac{\lambda^2}{(16\pi^2)^2}. \end{equation}
This coincides with the perturbative two loop result \cite{14}.
Insertion of (\ref{2leta}) into eq. (\ref{buteta}) yields for the
\(\beta_\lambda\)-function \begin{equation} \label{42}
\beta_\lambda = \frac{N+8}{16\pi^2}\lambda^2 -
\frac{9N+42}{(16\pi^2)^2}\lambda^3. \end{equation}  Again, the two
loop coefficient \cite{14} is exactly reproduced. Our "nested one
loop calculation" gives in next to leading order in \(\lambda\)
exactly the same results as a two loop calculation!
\setcounter{equation}{0} \section{Large quartic scalar coupling} In
the preceding sections we have concentrated on small values of
$\lambda$. Our nonperturbative exact evolution equations are by no
means restricted to this case. We demonstrate this here by
discussing the leading order contribution to $\beta_\lambda$ for
very large values of $\lambda$. We also combine approximately the
two regimes for large and small quartic couplings in order to
obtain the $\beta$-function for the whole range of $\lambda$. This
is justified by the existence of only a small transition regime
which is reached quickly for scales only somewhat below the
ultraviolet cutoff. Our starting point are the exact formulae
(\ref{betakappa}) and (\ref{betalambda}) for $\beta_\kappa$ and
$\beta_\lambda$ which we combine to \begin{eqnarray}
\label{combination} \beta_\lambda &=& (d-4+2\eta)\lambda
+2v_d(N-1)\lambda^2l_2^d \nonumber\\
&&+2v_d(N-1)\left\{2\lambda{z}_1l_2^{d+2}
+u_3{z_1}/{\lambda}l_1^{d+2}-z_2l_1^{d+2}+z_1^2l_2^{d+4}\right\}
\nonumber\\
&&+2v_dl_1^ds_1^d(2\lambda\kappa)\left\{2u_{3}^2\kappa/
\lambda-2u_3-2u_4\kappa\right\}\nonumber\\
&&+2v_dl_1^{d+2}s_1^{d+2}(2\lambda\kappa)\left\{\tilde{z}_1u_3/
\lambda-\tilde{z}_2\right\}\nonumber\\
&&+2v_d\bigg\{\left(3\lambda+2u_3\kappa\right)^2l_2^d
s_2^d(2\lambda\kappa)+2\left(3\lambda+2u_3\kappa\right)
\tilde{z}_1l_2^{d+2}s_2^{d+2}(2\lambda\kappa)\nonumber\\
&&+\tilde{z}_1^2l_2^{d+4}s_2^{d+4}(2\lambda\kappa)\bigg\}.
\end{eqnarray} Our strategy for an evaluation of $\beta_\lambda$
will consist of a series of assumptions on the maximal value (in
powers of $\lambda$) that different quantities on the r.h.s. of eq.
(\ref{combination}) can take for very large values of $\lambda$.
{}From there we compute the leading contribution to $\beta_\lambda$.
In a second step we establish the self consistency of our
assumptions by discussing explictly the evolution equations for the
quantities of concern. \\ Eq. (\ref{betakappa}) gives no indication
that $\kappa$ should be suppressed by inverse powers of $\lambda$
and we assume in the following $\kappa=O(1)$ in the sense of
counting powers of $\lambda^{-1}$. Then the threshold functions
$s_n^d(2\lambda\kappa)$ give suppression factors
$\sim\lambda^{-(n+1)}$. We will restrict the discussion of this
section to $N\ge 2$ where the contributions $\sim\lambda^{-(n+1)}$
are small compared to the Goldstone boson contributions $\sim
(N-1)$. Neglecting terms $\sim s_n^d(2\lambda\kappa)$ in a first
approximation we find \begin{eqnarray} \label{betalarge}
\beta_\lambda &=& 2v_d(N-1)\lambda^2l_2^d +(d-4+2\eta)\lambda
\nonumber\\ &&+2v_d(N-1)\left\{2\lambda{z}_1l_2^{d+2}
+u_3{z_1}/{\lambda}l_1^{d+2}-z_2l_1^{d+2}+z_1^2l_2^{d+4}\right\}.
\end{eqnarray} We remind that the quantities $l_n^d$ (\ref{smallL})
have to be evaluated here with $P = Z_k(\rho,q^2)q^2+R_k(q^2)$
according to (\ref{exprops}). The difference between this and the
values for small $\lambda$ (where $P\simeq Z_kq^2+R_k(q)$) as well
as the last term in (\ref{betalarge}) can be attributed to the wave
function renormalization $Z_k(\rho,Q^2)/Z_k$. We therefore need an
estimate of this quantity.\\ Let us start at some ultraviolet
cutoff $\Lambda$ with $Z_\Lambda(\rho,Q^2)$ depending only weakly
on $\rho$ such that $|Z'_\Lambda(\rho,Q^2)|\le O(1)$ and
$|Z''_\Lambda(\rho,Q^2)|\le O(\lambda)$. We first assume that this
situation remains valid for all scales $k$ for wich $\lambda$
remains large. Assuming also that $|u_3|$ is at most $O(\lambda^2)$
we find the simple leading order expression \begin{equation}
\label{betalarge2} \beta_\lambda = 2v_d(N-1)\lambda^2l_2^d
+(d-4+2\eta)\lambda.  \end{equation} Here the unspecified
$Q^2$-dependence of $Z_k(\rho,Q^2)$ only affects the value of
$l_2^d$ but not the general structure of eq. (\ref{betalarge2}).
Unless the  $Q^2$-dependence of $Z_k(\rho_0,Q^2)$ is extremely
dramatic $l_2^d$ will always remain of order one. As long as
$\frac{\partial}{\partial Q^2}\Gamma_k^{(4)}(Q,-Q,q,-q)$ does not
exceed a value $\sim O(1)$ and $\frac{\partial}{\partial
Q^2}\Gamma_k^{(3)}(Q,q,-q-Q)\sim O(1)$, $\frac{\partial}{\partial
Q^2}\Gamma_k^{(2)}(Q+q)\sim O(1)$ we conclude from the exact
relation (\ref{exzet}) that $\eta$ remains of the order $O(1)$.
With all these assumptions one finds the leading order contribution
\begin{equation} \label{betalarge3} \beta_\lambda =
2v_d(N-1)\lambda^2l_2^d.  \end{equation} Neglecting in addition the
$Q^2$-dependence of $Z_k(\rho_0,Q^2)$ the coefficients $l_2^d$ are
given in appendix C with $l_2^4=1$. Before a discussion of the
validity of our assumptions several comments are in order at this
place:\\ \vskip 0.3cm {\bf i)} For large $N$ we recover the result
of the leading contribution in the large $N$ approximation
\cite{17} \begin{eqnarray} \beta_\lambda &=& 2v_dN\lambda^2l_2^d
+(d-4)\lambda, \nonumber\\ \eta &=&0. \end{eqnarray} We remind that
the large $N$ approximation obtains formally by neglecting a
collective excitation which roughly corresponds to the radial mode.
With our assumptions the contributions from the radial mode are
small because of the threshold functions $s_n^d(2\lambda\kappa)$
and similar suppressions from the propagator of the massive radial
mode $\sim\lambda^{-1}$ in other expressions. This motivates the
qualitative correctness of (\ref{betalarge3}) also for moderate
values of $N$ (except $N=1$). \\ \vskip 0.3cm {\bf ii)} We note
that there is no dramatic difference in the behaviour of
$\beta_\lambda$ for small and large $\lambda$. We may therefore
extrapolate for intermediate values of $\lambda$ by connecting
$\beta_\lambda$ for small and large $\lambda$ at the transition
value $\lambda_{tr}$ where they take equal values. Roughly speaking
$\lambda_{tr}$ marks the transition from the perturbative to the
nonperturbative regime. We define \begin{equation}
\label{betatilde}
\tilde{\beta}=(\beta_\lambda-(d-4)\lambda)/(2v_dl_2^d\lambda^2)
\end{equation} such that, for $d=4$, \begin{equation} \label{cases}
\tilde{\beta}=\cases{N+8 -\frac{9N+42}{16\pi^2}\lambda &for
$\lambda<\lambda_{tr}$\cr N+\gamma&for $\lambda>\lambda_{tr}$\cr}.
\end{equation} (The following discussion will remain valid with
minor quantitative modifications in arbitrary dimension $d$. We
also allow for a correction factor $\gamma$ accounting for a
possible modification of the leading order result
(\ref{betalarge2}) which corresponds to $\gamma=-1$.) From
(\ref{cases}) we obtain the transition point \begin{equation}
\label{trans} \lambda_{tr}=\frac{16\pi^2(8-\gamma)}{9N+42}.
\end{equation} (For $N=4$, $\gamma=-1$ the transition point is
$\lambda_{tr}=18.2$.) At $\lambda_{tr}$ the ratio of the two loop
to one loop contribution for $\tilde{\beta}$ is given by
\begin{equation} \frac{\tilde{\beta}^{(2)}}{\tilde{\beta}^{(1)}} =
-\frac{8-\gamma}{N+8} \end{equation} which is of the order one
except for very large $N$.\\ \vskip 0.3cm {\bf iii)} The precise
form of $\tilde{\beta}$ for large $\lambda$, e.g. the value of
$\gamma$ as long as $\gamma$ is sufficiently large compared to
$-N$, is not very important for the qualitative behaviour of our
model. Taking into account the leading contribution
\begin{equation} \label{largegam} \frac{\partial\lambda}{\partial
t} = 2v_dl_2^d(N+\gamma)\lambda^2 \end{equation} one finds for the
scale dependence of $\lambda$ \begin{equation} \label{sol}
\frac{1}{\lambda(k)}=\frac{1}{\lambda(\Lambda)}
+2v_dl_2^d(N+\gamma)\ln\frac{\Lambda}{k}. \end{equation} For $d=4$
the value $\lambda_{tr}$ for the transition to the perturbative
behaviour is reached for $k>k_{tr}$  \begin{equation}
\label{ktrans}
k_{tr}=\Lambda\exp-\frac{9N+42}{(N+\gamma)(8-\gamma)}.
\end{equation} (The scale $k_{tr}$ is obtained for
$\lambda^{-1}(\Lambda)=0$. For positive values of
$\lambda^{-1}(\Lambda)$ the transition occurs at higher values of
$k$. This is used below to establish an upper bound on the Higgs
scalar mass.) For $N=4$ one finds \begin{eqnarray}
\frac{k_{tr}}{\Lambda}=\cases{0.056&for $\gamma=-1$\cr 0.087&for
$\gamma=0$\cr}. \end{eqnarray} Obviously the theory reaches the
perturbative regime very fast. In four dimensions there is no
$\varphi^4$-theory with large renormalized quartic coupling at
scales much smaller than the ultraviolet cutoff. For $k$ below the
scale $k_{pt}$ where perturbation theory becomes valid we use the
solution of (\ref{42}) with $\lambda_{pt}=\lambda(k_{pt})$
\begin{eqnarray} \label{bound} \frac{1}{\lambda(k)}
+\frac{9N+42}{16\pi^2(N+8)}\ln\left(\frac{16\pi^2(N+8)}{9N+42}
\frac{1}{\lambda(k)}
-1\right) \nonumber\\ =\frac{1}{\lambda_{pt}}
+\frac{9N+42}{16\pi^2(N+8)}\ln\left(\frac{16\pi^2(N+8)}{9N+42}
\frac{1}{\lambda_{pt}}
-1\right)+\frac{N+8}{16\pi^2}\ln\frac{k_{pt}}{k}\nonumber\\
=\frac{N+8}{16\pi^2}\ln\frac{\Lambda}{k}-
\frac{9N+42}{16\pi^2(N+\gamma)}+\frac{9N+42}{16\pi^2(N+8)}
\ln\frac{N+\gamma}{8-\gamma}.
\end{eqnarray} For the last equation we have inserted
$k_{pt}=k_{tr}$ (\ref{ktrans}). This constitutes an upper bound for
possible values $\lambda(k)$ for arbitrary scales $k<k_{tr}$. We
observe that this upper bound does not depend strongly on $\gamma$,
especially for large ratios $\Lambda/k$. Let us compare the result
(\ref{bound}) with the result of high order strong coupling
expressions on the lattice \cite{15}, namely \begin{equation}
\frac{1}{\lambda(k)}
-\frac{9N+42}{16\pi^2(N+8)}\ln\left(\frac{N+8}{16\pi^2}
\lambda(k)\right)
= \frac{N+8}{16\pi^2}\left(\ln\frac{\Lambda}{k}+C\right),
\end{equation} where $C=1.9$ for $N=4$. We observe agreement in the
large $N$ limit and also in leading order $\sim\lambda^{-1}$. The
constant $C$ computed from (\ref{bound}) obtains as $C=-2.8$ and
deviates from the lattice results. This may due to our neglection
of nonleading terms in the $\beta$-function for large $\lambda$ and
in an inaccurate matching between the strong coupling and weak
coupling regimes. Also the lattice definition of $\Lambda$ is not
identical with our definition. If we include in the strong coupling
$\beta$-function (\ref{betalarge3}) a nonleading term
$\delta\cdot\lambda$ and determine $\delta$ such that at
$\lambda_{tr}$ both $\beta_\lambda$ and
$\frac{\partial\beta_\lambda}{\partial\lambda}$ match for the
strong and weak coupling regimes, we find (for $\gamma=-1$) $\delta
=0.26, C=-1.8$. \vskip 0.3cm {\bf iv)} The upper bound on
$\lambda(k)$ translates into an upper bound for the scalar mass
both in the symmetric phase and in the phase with spontaneous
symmetry breaking. This bound depends on the ultraviolet cutoff
$\Lambda$. In the phase with spontaneous symmetry breaking the
value of the renormalized field corresponding to the potential
minimum goes to a constant $\rho_0$ larger than zero for $k\to 0$.
(The scaling solution with almost constant $\kappa$ for small
enough $\lambda$ ceases to be valid for $k$ smaller than some
characteristic scale of the order $k^2\simeq\lambda(k)\rho_0$.) The
renormalized mass is then approximately given\footnote{We define
the mass for nonvanishing momentum in order to avoid infrared
problems from the Goldstone bosons. Since $\lambda(k)$ changes
slowly for small enough $\lambda$ we can take $\lambda(\rho_0)$
instead of the appropriate four point function at nonvanishing
momentum with a good accuracy.} by \begin{equation} \label{trivial}
M^2=2\lambda(\rho_0)\rho_0. \end{equation} With
$\sqrt{2\rho_0}\simeq 246 GeV$ we obtain an upper bound $M_{max}$
as a function of $\Lambda$ by inserting the upper bound for
$\lambda(\rho_0)$ from eq. (\ref{bound}). It is qualitatively
similar to the lattice results \cite{15},\cite{16} and deviates
only in an overall factor of the constant $C$. \vskip 0.3cm  {\bf
v.)} One should note that $\beta_\lambda$ can be written as a
function of $\lambda$ on a given trajectory, i.e. for given initial
values of $\Gamma_\Lambda$. This is achieved by expressing all
other couplings on the r.h.s. of (\ref{combination}) in terms of
$\lambda$ by inserting the solution $\lambda(k)$, i.e.
$u_3(\lambda)=u_3(k(\lambda))$ etc., provided $\lambda(k)$ is a
monotonic function. We emphasize, however, that this function
$\beta_\lambda(\lambda)$ is {\em not} a universal function. In
general the details of $\beta_\lambda(\lambda)$ depend on the
particular initial values of the couplings specifying
$\Gamma_\Lambda$. Only in the vicinity of a fixpoint where
$\lambda$ varies only slowly, $\kappa$ is fixed by its
($\lambda$-dependent) critical value corresponding to the critical
trajectory, and all other couplings take their values at infrared
fixpoints as functions of $\lambda$, the function
$\beta_\lambda(\lambda)$ becomes universal. In four dimensions this
happens for small $\lambda$ (the vicinity of the Gaussian
fixpoint). In contrast, for very large values of $\lambda$ the
evolution is extremely rapid and one is far away from any fixpoint
behaviour. Details of $\beta_\lambda(\lambda)$  for large $\lambda$
are therefore indeed expected to depend on detailes of the short
distance action $\Gamma_\Lambda$. The problem of estimating
$\beta_\lambda(\lambda)$ for the most general form of
$\Gamma_\Lambda$ is certainly not solvable. In extreme cases
$\Gamma_\Lambda$ may correspoond to a theory in a universality
class completely different from the $\varphi^4$-theory. For
example, the effective action $\Gamma_\Lambda$ for a theory with
light fermions and Yukawa couplings to the scalars can be written
in a form of a (nonlocal) action involving only scalar fields by
integrating out the fermionic degrees of freedom. This would
certainly lead to a different function $\beta_\lambda(\lambda)$
even for small values of $\lambda$. (Additional contributions from
Yukawa couplings would arise.) We will be satisfied to establish
that the qualitative behaviour of (\ref{cases}) holds for a large
class of initial values $\Gamma_\Lambda$ within the range of
attraction of the Gaussian fixpoint. The discussion of the validity
of the assumptions leading to (\ref{cases}) will also give
information on this range of attraction. For initial values in the
vicinity of the boundary of the range of attraction  of the
Gaussian fixpoint we expect qualitatively different evolution. No
"triviality bound" of the type discussed before can be derived in
this case. \\ \vskip 0.3cm  Let us now discuss the assumptions
leading to (\ref{cases}) in more detail. The most important one was
that $\kappa$ is not of the order $\lambda^{-1}$ or smaller such
that contributions from the radial mode $\sim
s_n^d(2\lambda\kappa)$ can be neglected. For
$\kappa\sim\lambda^{-1}$ the leading contribution to the evolution
of $\kappa$ would be given by (\ref{betakappa}) \begin{equation}
\frac{\partial\kappa}{\partial t} =
\frac{N-1}{16\pi^2}\left(1+\frac{z_1}{\lambda}\right).
\end{equation} This would drive $\kappa$ very quickly to zero and
we conclude that values $\kappa\sim\lambda^{-1}$ or smaller
correspond to the symmetric phase. We are interested in the phase
with spontaneous symmetry breaking near the phase transition. Here
$\kappa$ cannot be smaller than of order one. Our next worry is the
value of the dimensionless $\varphi^6$-coupling $u_3$. If $u_3$
would be of the order $\lambda^3$ the suppression due to the
threshold functions is insufficient and $\beta_\lambda$ would
obtain contributions $\sim\lambda^3$. The validity of the
approximation (\ref{betalarge}) requires that $u_3$ is at most of
the order $\lambda^2$. At first sight this seems to contradict the
one loop estimate (\ref{u3loop}). We should remember, however, that
large values of $\lambda$ correspond to a situation with rapidly
varying couplings, in contrast to the approximative fixpoint
behaviour for small and slowly varying $\lambda$. For an estimate
of $u_3$ we integrate its leading order evolution equation which
may be obtained from eq. (\ref{four}) after derivation with respect
to $\rho$ and introduction of the dimensionless renormalized
couplings (\ref{dimren}) \begin{equation} \label{largeu}
\frac{\partial u_3}{\partial t}= -4v_dl_3^d(N-1)\lambda^3 +
6v_dl_2^d(N-1)\lambda{u}_3. \end{equation}  As long as $u_3$ is
small compared to $\lambda^2$ we can use the solution (\ref{sol}).
The first term in (\ref{largeu}) leads to an increase in $u_3$
until it reaches a value $\sim\lambda^2$. Then the second term in
(\ref{largeu}) becomes important and $u_3/\lambda^2$ is driven to a
zero of the r.h.s. of \begin{equation} \frac{\partial}{\partial
t}\frac{u_3}{\lambda^2}=\lambda^{-2}\frac{\partial u_3}{\partial t}
-2\lambda^{-3}\frac{\partial \lambda}{\partial t}=
\frac{N-1}{16\pi^2}\lambda\left(-2l_3^4+l_2^4
\frac{u_3}{\lambda^2}\right)
\end{equation} (for $d=4, \gamma=-1$). We note that the fixpoint
\begin{equation} \label{lufix}
\left(\frac{u_3}{\lambda^2}\right)_\star=2\frac{l_3^4}{l_2^4}
\end{equation} corresponds now to a value of $u_3$ of the order
$\lambda^2$. The difference compared to (\ref{u3fix}) arises from
the second term in (\ref{largeu}) which has been neglected in sect.
4. By similar arguments it is easy to establish that $u_4$ is at
most of the order $\lambda^3$ in agreement with our
approximations.\\ Let us finally make an estimate for the moments
$z_z$ and $z_2$. The one loop estimate (\ref{zmoments}) suggests
values $z_1\sim\lambda^2$, $z_2\sim\lambda^3$ in contradiction to
the validity of our approximations. If we write the evolution
equations for $z_i$ in the form \begin{equation}
\frac{\partial}{\partial t}z_i=
-A_i(\lambda)+B_i(\lambda)z_i+\ldots \end{equation} only the terms
$A_1\sim\frac{\lambda^2}{16\pi^2},
A_2\sim\frac{\lambda^3}{16\pi^2}$ are directly related to the one
loop formulae of sect. 5. (For the discussion of section 5 it is
implicitely assumed that $B_i$ is dominated by the canonical
dimension of $Z', Z''$. This is only valid for small $\lambda$.) In
leading order the evolution equation for $Z'_k(\rho_0,Q^2=0)$ and
$Z''_k(\rho_0,Q^2=0)$ (\ref{azeta}), (\ref{bzeta}) yields
\begin{eqnarray} \label{largez1} \frac{\partial}{\partial
t}(z_1\kappa)&=& 2v_d\lambda\left(\frac{4}{d}m_3^d +
(N-1)l_2^dz_1\kappa\right), \\ \label{largez2}
\frac{\partial}{\partial t}(z_2\kappa)&=&
4v_d(N-1)l_2^dz_2\kappa\lambda\nonumber\\
&&-4v_d\lambda^2\left(\frac{6}{d}m_4^d+(N-1)l_3^dz_1\kappa\right)
\nonumber
\\&&+2v_du_3\left(\frac{4}{d}m_3^d + (N-1)l_2^dz_1\kappa\right).
\end{eqnarray} Here we have approximated $L_{1,1}^d(w_1,w_2)\simeq
\frac{1}{w_2}L_{1,0}^d(w_1)$ and similar for other threshold
functions and we have omitted terms $\sim Y_k$. (This becomes exact
in the large $N$ limit). We conclude that for large $\lambda$ there
are infrared fixpoints in $z_1\kappa$ and the ratio
$z_2\kappa/\lambda$ \begin{eqnarray}
(z_1\kappa)_\star&=&-\frac{4m_3^d}{(N-1)dl_2^d}=-\frac{1}{N-1},\\
\left(\frac{z_2\kappa}{\lambda}\right)_\star&=&
\frac{2}{(N-1)dl_2^d}\left(3m_4^d-2\frac{l_3^d}{l^d_2}m_3^d\right)
=-\frac{1}{N-1}\left(3\ln\frac{4}{3}-\frac{3}{4}\right)
\end{eqnarray} (Here we have inserted the fixpoint value for $u_3$
(\ref{lufix}) and the last part of the equations is evaluated for
$d=4$ with constants $l_n^4,m_n^4$ given in appendix C.) From this
we infer that $|z_1|$ grows within the regime where $\lambda$ is
large at most to a value $\sim O(1)$ and $|z_2|$ becomes at most
$\sim O(\lambda)$. In consequence the last term in
(\ref{betalarge}) can indeed be neglected in leading order in the
regime of very strong quartic scalar coupling. With the estimations
\begin{eqnarray} \label{estim1} |\kappa|\le O(1)\quad&,&|u_3|\le
O(\lambda^2)\quad,\quad |u_4|\le O(\lambda^3),\nonumber\\ |z_1|\le
O(1)\quad&,&\quad |z_2|\le O(\lambda) \end{eqnarray} and
$s_n^d(2\lambda\kappa)\sim O(\lambda^{-(n+1)})$ we find that the
approximation (\ref{betalarge2}) is indeed valid provided
\begin{eqnarray} \label{estim2} |\tilde{z}_1|\le
O(\lambda^2)\quad,\quad |\tilde{z}_2|\le O(\lambda^3).
\end{eqnarray} We therefore quote the evolution equation for
$\tilde{Z}_k$ wich has been computed in sect. 5 of \cite{4} and
reads in leading order for large $\lambda$ \ba
\frac{\partial\tilde{Z}_k(\rho)}{\partial
t}&=&v_dk^{d-2}(N-1)Y'_k(\rho)\rho L_{1,0}^d(U'_k(\rho))
+\frac{4v_d}{d}k^{d-6}(N-1){U''_k}^2(\rho)\rho
M^d_{4,0}(U'_k(\rho))\nonumber\\
&&-2v_dk^{d-4}(N-1)Y_k(\rho)U''_k(\rho)\rho L_{2,0}^d(U'_k(\rho)).
\ea Expressed in dimensionless renormalized couplings we obtain for
the scale dependence of $\tilde{Z}$ and its derivatives in leading
order ($\tilde{z}_0\equiv \tilde{Z}/Z$) \ba
\frac{\partial\tilde{z}_0}{\partial t}&=&-2v_d(N-1)l_1^dy_1\kappa
-\frac{8v_d}{d}(N-1)m^d_4\lambda^2\kappa \nonumber\\
&&+4v_d(N-1)l_2^dy_0\lambda\kappa, \ea \ba
\frac{\partial\tilde{z}_1(\rho)}{\partial
t}&=&-2v_d(N-1)l_1^dy_2\kappa
-\frac{16v_d}{d}(N-1)m^d_4u_3\lambda\kappa \nonumber\\
&&+4v_d(N-1)l_2^dyu_3\kappa+6v_d(N-1)l_2^dy_1\kappa\lambda
\nonumber\\
&&+\frac{32v_d}{d}(N-1)m_5^d\lambda^3\kappa-8v_d(N-1)l_3^d
y_0\lambda^2\kappa
\ea and \ba \frac{\partial\tilde{z}_2(\rho)}{\partial
t}&=&-2v_d(N-1)l_1^dy_3\kappa+8v_d(N-1)l_2^dy_2\kappa\lambda
\nonumber\\&&+10v_d(N-1)l_2^dy_1\kappa
u_3
-\frac{16v_d}{d}(N-1)m^d_4\left(u_3^2+u_4\lambda\right)\kappa
\nonumber\\
&&+\frac{160v_d}{d}(N-1)m_5^du_3\lambda^2\kappa+
4v_d(N-1)l_2^du_4y_0\kappa\nonumber\\
&&-20v_d(N-1)l_3^dy_1\kappa\lambda^2-
24v_d(N-1)l_3^du_3y_0\kappa\lambda\nonumber\\
&&-\frac{160v_d}{d}(N-1)m_6^d\lambda^4\kappa+
24v_d(N-1)l_4^dy_0\kappa\lambda^3.
\ea We recognize that ($y_0=k^{d-2}Z^{-2}Y$) \be
\tilde{z}_n=\kappa{y}_n \ee holds in leading order and that the
system above exhibits fixpoints of the orders \ba
(y_n)_\star\sim\lambda^{n+1}. \ea These results agree with our
assumptions made above. \\ The only point missing for a proof that
eq. (\ref{cases}) is indeed the leading contribution for large
$\lambda$ concerns the neglection of the $Q^2$-dependence of
$Z_k(\rho,Q^2)$ and $\eta$. If this momentum dependence contributes
only in nonleading order in $\lambda$ one obtains $\gamma=-1$ for
$d=4$. It is also conceivable, however, that the momentum
dependence modifies the value of $\gamma$ and this is the reason
why we have kept $\gamma$ as a free parameter at the present stage.
In principle, these questions can be answered by establishing an
appropriate fixpoint behaviour but we have not performed this task
in the present paper. It may even be hoped that rigorous bounds on
the maximal size of the various quantities appearing on the r.h.s.
of (\ref{combination}) could be proven with the use of exact
evolution equations, in a spirit similar to our discussion for
$u_3,z_1$ and $z_2$. Even without going so far we believe that our
estimate (\ref{cases}) for $\beta_\lambda$ is reliable for very
large values of $\lambda$, with a certain uncertainity in the
region of intermediate $\lambda$. (There is disagreement about the
behaviour of $\beta_\lambda$ for large $\lambda$ in the literature
\cite{18}. These references find the asymptotic behaviour
$\beta_\lambda\sim\lambda, \beta_\lambda\sim\lambda^2$ and
$\beta_\lambda\sim\lambda^{\frac{3}{2}}$.) Excluding
$\tilde{\beta}$ to be almost zero for intermediate $\lambda$ this
behaviour is sufficient to derive the "triviality bound"
(\ref{trivial}) for the mass of the Higgs scalar. We emphasize that
our treatment establishes the existence of a triviality bound
without the use of a lattice regularization. For example, we could
use a smooth ultraviolett cutoff $\Lambda$ within our formulation
in continuous space. This determines the "initial value"
$\Gamma_\Lambda$ for a given bare theory. Triviality bounds appear
for a wide class of bare actions for wich $\Gamma_\Lambda$ obeys
(\ref{estim1}) and (\ref{estim2}) and similar restrictions for the
momentum dependence of the vertices in $\Gamma_\Lambda$.
\setcounter{equation}{0} \section{Conclusion} We have computed the
\(\beta\)-function in order \(\sim\lambda^3\) and the anomalous
dimension in order \(\sim\lambda^2\) without ever doing a two loop
calculation. We used instead an exact evolution equation which has
a one loop form, with classical propagator replaced by the full
propagator and solved it using a nonperturbative method which
combines exact evolution equations for "independent" couplings with
renormalization group improved one loop expressions for "secondary"
ones. It is impressive to see how different properties of the
propagator in a background field combine to reproduce the
perturbative two loop result in next to leading order in the small
quartic coupling. We emphasize that in the course of our
calculation we have in addition gained a lot of information on the
momentum dependence of various 1PI Green functions. In contrast to
the more formal perturbative loop expansion the different
contributions to the anomalous dimension and the \(\beta\)-function
for the quartic coupling are here directly connected with physical
properties, i.e. the behaviour of Green functions.\\ Even though
the reproduction of perturbative two loop results by a one loop
calculation is interesting by its own, this was not the main
motivation of the present work. The two loop results are known
since a long time and the present method does not offer
calculational simplification in this respect. The main interest
lies in the fact that our method is by no means restricted to an
expansion in a small coupling \(\lambda\). The exact evolution
equation is valid for arbitrary values of the quartic coupling. We
have demonstrated this by computing the $\beta$-function also for
very large values of $\lambda$. Our results confirm the lattice
results for the triviality bound for the Higgs scalar mass in the
context of a Euclidean field theory formulated in continuous
space.\\ A computational method giving reliable results also for
large $\lambda$ is particulary important in dimensions less than
four, where it is well known that the critical behaviour is
governed by a large value of the infrared fixpoint for \(\lambda\).
Our method can be directly applied to arbitrary dimensions and
arbitrary values of the coupling. The main modification as compared
to the present calculation is the size of the dimensionless mass
term \(2\lambda\kappa\) in the propagator. It cannot be treated as
a small quantity anymore and the threshold functions have to be
evaluated numerically. In view of the nonperturbative applications
it is crucial to have a consistent nonperturbative scheme for an
approximative solution of the exact evolution equation. This was
presented in the present paper and amounts to a renormalization
group improved one loop calculation of the "secondary" couplings.
The successful precision test of our systematic truncation scheme
offers the exciting prospect of precision calculations for the
critical behaviour in two and three dimensions or the high
temperature phase transition in four dimensions.
\setcounter{equation}{0}  \newpage \appendix \section{One loop
calculation of the wave function renormalization} In this section
we perform a one loop calculation of the wave function
renormalization. From eq. (\ref{expan}) we get after functional
derivation and inserting a configuration with constant scalar field
\(\varphi_1=\sqrt{2\rho}\), \(\varphi_a=0\) for \(a\not= 1\)
\begin{eqnarray} \left(\Gamma_k^{(2)}(Q,-Q;\rho)\right)_{ab} &=&
U'_k(\rho)\delta_{ab}+2\rho{U}''_k(\rho)\delta_{a1}\delta_{b1}
\nonumber\\
&&+Q^2\left[Z_k(\rho,Q^2)\delta_{ab}+\rho{Y}_k(\rho,Q^2)
\delta_{a1}\delta_{b1}\right].
\end{eqnarray} From this we see \begin{eqnarray} \label{9.2}
Q^2{Z}_k(\rho,Q^2) &=&
\left(\Gamma_k^{(2)}(Q,-Q;\rho)\right)_{22}-
\left(\Gamma_k^{(2)}(0,0;\rho)\right)_{22},\nonumber\\
Q^2\tilde{Z}_k(\rho,Q^2) &=&
\left(\Gamma_k^{(2)}(Q,-Q;\rho)\right)_{11}-
\left(\Gamma_k^{(2)}(0,0;\rho)\right)_{11}
\end{eqnarray} and \begin{eqnarray} Z_k(\rho,Q^2=0) &=&
\lim_{Q^2\to{0}}\frac{\partial}{\partial{Q^2}}
\left(\Gamma_k^{(2)}(Q,-Q;\rho)\right)_{22},\nonumber\\
\tilde{Z}_k(\rho,Q^2=0) &=&
\lim_{Q^2\to{0}}\frac{\partial}{\partial{Q^2}}
\left(\Gamma_k^{(2)}(Q,-Q;\rho)\right)_{11}.
\end{eqnarray} We consider the difference \begin{eqnarray}
Z_k(\rho,Q^2)-Z_k(0,0) =
\frac{1}{Q^2}\left[\left(\Gamma_k^{(2)}(Q,-Q;\rho)\right)_{22}
-\left(\Gamma_k^{(2)}(0,0;\rho)\right)_{22}\right]-
\frac{\partial}{\partial{Q^2}}
\left(\Gamma_k^{(2)}(Q,-Q;0)\right)_{22}\bigg\vert_{Q^2=0}
\end{eqnarray} as one of the "secondary" couplings
\(\Gamma_k^{(n)(L)}\) in the sense of section 2 and similar for
\(\tilde{Z}_k\). (Note \(\tilde{Z}_k(0,0)=Z_k(0,0)\).) In order to
evaluate the one loop formula (\ref{split}) we approximate
\(\Gamma_k^{(2)}\) on the r.h.s. by the second functional
derivative of the expression \begin{equation}
\Gamma_{k}[\varphi]=\int{d}^{d}{x}\bigg\{{U}_{k}(\rho)+
\frac{1}{2}Z_k\partial_\mu\varphi^{a}\partial^\mu\varphi_{a}\bigg\}.
\end{equation} This yields  \begin{eqnarray} \label{loopq}
Z_k(\rho,Q^2) &=& Z_k(0,0) -
2\frac{{U''_k}^2(\rho)\rho}{Q^2}(2\pi)^{-d}\int{d}^{d}q
\left(P(q)+U'_k(\rho)+2U''_k(\rho)\rho\right)^{-1}
\nonumber\\
&&\cdot\left[(P(q+Q)+U'_k(\rho))^{-1}-(P(q)+U'_k(\rho))^{-1}\right],
\nonumber\\
\tilde{Z}_k(\rho,Q^2) &=& {Z}_k(0,0) -
\frac{\rho}{Q^2}(2\pi)^{-d}\int{d}^{d}q\Bigg\{
\frac{\left(3U''_k(\rho)
+2U'''_k(\rho)\rho\right)^2}{P(q)+U'_k(\rho)+2U''_k(\rho)\rho}
\nonumber\\
&&\cdot\left[\left(P(q+Q)+U'_k(\rho)+2U''_k(\rho)\rho\right)^{-1}
-\left(P(q)+U'_k(\rho)+2U''_k(\rho)\rho\right)^{-1}\right]
\nonumber\\ &&
+\frac{(N-1){U_k''}^2(\rho)}{P(q)+U'_k(\rho)}
\left[\left(P(q+Q)+U'_k(\rho)\right)^{-1}-
\left(P(q)+U'_k(\rho)\right)^{-1}\right]\Bigg\}
\end{eqnarray} and \begin{eqnarray} \label{loopzero} Z_k(\rho,0)
&=& {Z}_k(0,0)
+8\frac{v_d}{d}{U_k''}^2\rho\int_{0}^\infty{d}xx^{\frac{d}{2}}
\dot{P}^2\left(P+U'_k\right)^{-2}\left(P+U'_k+2U''_k\rho\right)^{-2},
\nonumber\\ \tilde{Z}_k(\rho,0) &=& {Z}_k(0,0) +
4\frac{v_d}{d}(3U''_k+2U'''_k\rho)^2\rho
\int_{0}^\infty{d}xx^{\frac{d}{2}}
\dot{P}^2\left(P+U'_k+2U''_k\rho\right)^{-4}\nonumber\\
&&+4(N-1)\frac{v_d}{d}{U_k''}^2\rho
\int_{0}^\infty{d}xx^{\frac{d}{2}}\dot{P}^2\left(P+U'_k\right)^{-4}
\end{eqnarray} where
\(x=q^2\),\(\dot{P}=\frac{\partial{P}}{\partial{x}}\) and \(P\) is
given by (\ref{propfix}). The one loop expressions eqs.
(\ref{loopq}) and eqs. (\ref{loopzero}) are ultraviolet finite in
any dimension less than six. The renormalization group improved one
loop approximation should therefore only be used for \(d<6\).
\setcounter{equation}{0} \section{Evaluation of some integrals} In
this section we evaluate the integrals \(I(Q^2)\) and \(J(Q^2)\)
appearing in section 5 in four dimensions. Throughout this section
we will use the inverse propagator (\ref{propfix}) for
computations, i.e. we will set \(Z = 1\). We start evaluating the
\(I(Q^2)\) which is defined in eq. (\ref{Iinteg}) and reads, with
\(a=k^2\alpha\), \(b=k^2\beta\) in eq. (\ref{Iinteg})
\begin{eqnarray} I(Q^2) &\equiv &
\frac{1}{2}(2\pi)^{-4}\int{d}^{4}{q}P^{-1}(q)
\left[P^{-1}(q+Q)-P^{-1}(q)\right]
\nonumber\\ &=&
v_{4}\int\limits_0^{1}\int\limits_0^{1}{d}{a}{d}{b}
\left(e^{-\frac{Q^2}{k^2}\frac{ab}{a+b}}-1\right)(a+b)^{-2}.
\end{eqnarray} We transform
\(y=\frac{b^2}{a+b},dy=-\frac{y^2}{b^2}da\) and perform the
\(y\)-integration  \begin{equation}
I(Q^2)=v_4\frac{k^2}{Q^2}\int\limits_0^1{db}b^{-2}
\left(1-e^{-\frac{Q^2}{k^2}\frac{b}{1+b}}-\frac{Q^2}{k^2}b\right)
+v_4\ln{2}.
\end{equation} In terms of \(z=\frac{b}{1+b}\) one finds
\begin{equation} I(Q^2)=
v_4\frac{k^2}{Q^2}\int\limits_0^\frac{1}{2}dz\left(z^{-2}-
\frac{Q^2}{k^2}z^{-1}-z^{-2}e^{-\frac{Q^2}{k^2}z}\right).
\end{equation} Writing the integrand as a Taylor series one finally
obtains \begin{equation} I(Q^2)=
v_4\sum_{n=0}^\infty\frac{(-1)^{n+1}}{(n+2)!(n+1)}
\left(\frac{1}{2}\right)^{n+1}\left(\frac{Q^2}{k^2}\right)^{n+1}
\end{equation} which is eq. (\ref{Iintegral}). Analogously we
proceed with the \(J(Q^2)\)-integral. It reads \begin{eqnarray}
J(Q^2) &\equiv &
\frac{1}{2}(2\pi)^{-4}\int{d}^{4}{q}P^{-2}(q)
\left[P^{-1}(q+Q)-P^{-1}(q)\right]
\nonumber\\ &=&
v_4{k}^{-2}\int\limits_0^1{dc}\left[\ln(2+c)-2\ln(1+c)+\ln{c}+
\frac{k^2}{Q^2}\int\limits_\frac{c}{1+c}^\frac{1+c}{2+c}dz{z}^{-2}
\left(1-e^{-\frac{Q^2}{k^2}z}\right)\right].
\end{eqnarray} We consider the double integral \(K\) and integrate
by parts \begin{eqnarray}
K&\equiv&\frac{k^2}{Q^2}\int\limits_0^1{dc}
\int\limits_\frac{c}{1+c}^\frac{1+c}{2+c}dz{z}^{-2}
\left(1-e^{-\frac{Q^2}{k^2}z}\right)\nonumber\\
&=&\frac{k^2}{Q^2}\left\{-\ln2+\int\limits_0^1{dc}
\left[c^{-1}\left(1-e^{-\frac{Q^2}{k^2}\frac{c}{1+c}}\right)-
e^{\frac{Q^2}{k^2}\frac{c}{1+c}}+
\frac{2+c}{1+c}e^{\frac{Q^2}{k^2}\frac{1+c}{2+c}}+
\frac{Q^2}{k^2}\int\limits_\frac{c}{1+c}^\frac{1+c}{2+c}
\frac{dz}{z}e^{-\frac{Q^2}{k^2}z}\right]\right\}.\nonumber
\end{eqnarray} The brackets \([\cdots]\) contain four terms. In the
first one we substitute \(z=\frac{c}{1+c}\) and expand in a Taylor
series \begin{equation}
\frac{k^2}{Q^2}\int\limits_0^1{dc}c^{-1}
\left(1-e^{-\frac{Q^2}{k^2}\frac{c}{1+c}}\right)=
\sum_{n=0}^\infty\frac{(-1)^n}{(n+1)!}\left(\frac{Q^2}{k^2}\right)^n
\int\limits_0^\frac{1}{2}dz\frac{z^n}{1-z}.
\end{equation}  Similarly we proceed with the next two terms
\begin{eqnarray}
\frac{k^2}{Q^2}\int\limits_0^1{dc}e^{-\frac{Q^2}{k^2}\frac{c}{1+c}}
&=&\sum_{n=0}^\infty\frac{(-1)^{n+1}}{n!}
\left(\frac{Q^2}{k^2}\right)^{n-1}
\int\limits_0^\frac{1}{2}dz\frac{z^n}{(1-z)^2},\\
\frac{k^2}{Q^2}\int\limits_0^1{dc}
\frac{2+c}{1+c}e^{-\frac{Q^2}{k^2}\frac{1+c}{2+c}}
&=&\sum_{n=0}^\infty\frac{(-1)^{n}}{n!}
\left(\frac{Q^2}{k^2}\right)^{n-1}
\int\limits_\frac{1}{2}^\frac{2}{3}dz\frac{z^n}{(1-z)^2}.
\end{eqnarray} In the last term (the double integral) we expand the
integrand in a Taylor series and perform the
\(\int\limits_\frac{c}{1+c}^\frac{1+c}{2+c}dz\)-integration for
every term. Again we substitute \(z=\frac{c}{1+c}\) and
\(z=\frac{1+c}{2+c}\) \begin{equation}
\int\limits_0^1{dc}\int\limits_\frac{c}{1+c}^\frac{1+c}{2+c}
\frac{dz}{z}e^{-\frac{Q^2}{k^2}z}=6\ln2-3\ln3+
\sum_{n=0}^\infty\frac{(-1)^{n+1}}{(n+1)!(n+1)}
\left(\frac{Q^2}{k^2}\right)^{n+1}\left\{\int_0^\frac{1}{2}dz
\frac{z^{n+1}}{(1-z)^2}
-
\int_{\frac{1}{2}}^{\frac{2}{3}}dz\frac{z^{n+1}}{(1-z)^2}\right\}.
\end{equation} Summing up these contributions, one finally obtains
\begin{equation} J(Q^2) =
v_{4}k^{-2}\sum_{n=0}^\infty\frac{(-1)^n}{(n+2)!(n+1)}
\left(\frac{Q^2}{k^2}\right)^{n+1}\left\{\int_0^\frac{1}{2}dz
\frac{z^{n+1}}{(1-z)^2}
- \int_{\frac{1}{2}}^{\frac{2}{3}}dz\frac{z^{n+1}}{(1-z)^2}\right\}
\end{equation} which is eq. (\ref{Jintegral}).\\
\setcounter{equation}{0} \section{Contributions to
\(\beta_\lambda\)} Finally we add up all the contributions wich
enter in the \(\beta_\lambda\)-function (\ref{betalambdaref}) in
order \(\lambda^3\). Inserting eqs. (\ref{zmoments}),
(\ref{missdelta}) and (\ref{deltaeta}) in (\ref{betalambdaref}) we
write \begin{eqnarray} \label{betafin} \beta_\lambda &=&
2v_4(N+8)\lambda^2
-4v_4^2\lambda^3\Bigg\{(10N+44)\Bigg[2l_1^4{l}_3^4-\frac{1}{2}
\sum_{n=0}^\infty\frac{(-1)^n}{(n+2)!(n+1)}
\left(\frac{1}{2}\right)^{n}l_2^{6+2n}\nonumber\\
&&+2\sum_{n=0}^\infty\frac{(-1)^n}{(n+2)!(n+1)}l_1^{6+2n}
\left(\int_0^\frac{1}{2}dz\frac{z^{n+1}}{(1-z)^2}
-
\int_{\frac{1}{2}}^{\frac{2}{3}}dz\frac{z^{n+1}}{(1-z)^2}
\right)\Bigg]
\nonumber\\
&&-(N+2)\left(3m_4^4{l}_1^4-\frac{1}{2}\right)\Bigg\}.
\end{eqnarray} From eqs. (\ref{five}) and (\ref{smallL}) we find
\begin{equation}
l_n^d=n\int\limits_0^\infty{d}yy^{\frac{d}{2}-n}e^{-y}
\left(1-e^{-y}\right)^{n-1}
\end{equation} where \(y=x/k^2\) and \(f_k\) from eq.
(\ref{ffunc}). These integrals may be integrated by parts and lead
to \(\Gamma\)-functions. From this we get \begin{eqnarray}
l_1^4&=&l_2^4=1, \nonumber\\ l_3^4&=&3\ln\frac{4}{3},\nonumber\\
l_1^{6+2n}&=&(n+2)!, \nonumber\\
l_2^{6+2n}&=&2(n+1)!-\frac{(n+1)!}{2^{n+1}}.  \end{eqnarray} From
eqs. (\ref{mint}) and (\ref{smallM}) we obtain for \(n\ge
\frac{d}{2}+1\) with
\(\partial_t\left(\frac{x^n}{P^n}\dot{P}^2\right) =
-2x\partial_x\left(\frac{x^n}{P^n}\dot{P}^2\right)\) and
integration by parts \begin{eqnarray} m_n^d&=&
\delta_{n,\frac{d}{2}+1}+(n-1-\frac{d}{2})\int\limits_0^\infty
{d}yy^\frac{d}{2}\left(\frac{\partial{p}}{\partial{y}}\right)^{2}
p^{-n}\nonumber\\
&=&\delta_{n,\frac{d}{2}+1}+(n-1-\frac{d}{2})\epsilon_n^d
\end{eqnarray} with \begin{equation}
p(y)=\frac{P}{k^2}=\frac{y}{1-e^{-y}}. \end{equation} Directly we
find \begin{equation} m_3^4=1  \end{equation} and after further
integration by parts  \begin{equation} m_4^4=\frac{1}{2}.
\end{equation} (Note that the \(\epsilon_n^d\) are closely related
to the $m_n^d$.) Inserting these results in eq. (\ref{betafin}) it
remains performing two sums, namely \begin{eqnarray}
\sum_{n=0}^\infty\frac{(-1)^n}{(n+1)(n+2)}x^n
&=&\sum_{n=0}^\infty\frac{(-1)^n}{n+1}x^n -
\sum_{n=0}^\infty\frac{(-1)^n}{n+2}x^n \nonumber\\ &=&
x^{-1}\ln(1+x) + x^{-2}\left[\ln(1+x)-x\right] \end{eqnarray} with
\(x=\frac{1}{2}, x=\frac{1}{4}\) and \begin{eqnarray}
\sum_{n=0}^\infty\frac{(-1)^n}{n+1}\int_a^b{dz}
\frac{z^{n+1}}{(1-z)^2}
&=&
\int_a^b\frac{dz}{(1-z)^2}\sum_{n=1}^\infty\frac{(-1)^{n+1}}{n}z^n
\nonumber\\
&=&\left[\frac{\ln(1+z)}{1-z}-\frac{1}{2}\ln(1+z)+
\frac{1}{2}\ln(1-z)\right]\bigg\vert_a^b.
\end{eqnarray} Putting all this together leads to the desired
result.  \vskip 2cm {\Large\bf Figure Caption} \vskip
0.5cm {\bf Fig. 1:} Graphical representation of the exact evolution
equation \end{document}